\documentclass[12pt]{article}
\pdfoutput=1

\usepackage{amsmath,amssymb,amsthm,graphicx,amscd,multirow,xspace}
\usepackage[colorlinks=true,urlcolor=blue,anchorcolor=blue,citecolor=blue,filecolor=blue,linkcolor=blue,menucolor=blue,pagecolor=blue]{hyperref}
\usepackage[compress,numbers]{natbib}
\usepackage{subfigure, placeins}
\usepackage[all]{xy}
\usepackage{color}
\usepackage[usenames,dvipsnames,svgnames,table]{xcolor}
\usepackage{multirow}
\usepackage{bm}
\usepackage{dsfont}
\usepackage{mathrsfs}
\usepackage{bbold}
\usepackage{floatrow}
\newfloatcommand{capbtabbox}{table}[][\FBwidth]
\usepackage{arydshln}

\long\def\symbolfootnote[#1]#2{\begingroup%
\def\thefootnote{\fnsymbol{footnote}}\footnote[#1]{#2}\endgroup}

\newcommand{\newc}{\newcommand}
\newc{\gsim}{\lower.7ex\hbox{$\;\stackrel{\textstyle>}{\sim}\;$}}
\newc{\lsim}{\lower.7ex\hbox{$\;\stackrel{\textstyle<}{\sim}\;$}}
\newc{\gev}{\,{\rm GeV}}
\newc{\mev}{\,{\rm MeV}}
\newc{\ev}{\,{\rm eV}}
\newc{\kev}{\,{\rm keV}}
\newc{\tev}{\,{\rm TeV}}

\newcommand{\Ldash}[1]{\multicolumn{1}{|c}{#1}}
\newcommand{\Rdash}[1]{\multicolumn{1}{c|}{#1}}
\newcommand{\LRdash}[1]{\multicolumn{1}{|c|}{#1}}

\def\CS{{\mathcal{S}}}
\newc{\mz}{M_Z}
\newc{\mpl}{M_*}
\newc{\mw}{m_{\rm weak}}
\newc{\nr}[1]{N^c_R{}_{#1}}
\usepackage{amsmath}

\newcommand{\G}{\mathcal{G}}

\newcommand{\one}{\mathbb{1}}
\def\be{\begin{equation}}
\def\ee{\end{equation}}
\def\n{{\mathfrak{n}}}
\def\a{{\mathfrak{a}}}
\def\b{{\mathfrak{b}}}

\def\m{{\mathfrak{m}}}
\def\n{{\mathfrak{n}}}

\def\m{{\mathfrak{m}}}

\def\CC{{\mathcal{C}}}
\def\IZ{{\mathbb{Z}}}

\def\IR{{\mathbb{R}}}

\def\six{{{\bf 6}}}

\def\four{{{\bf 4}}}
\def\two{{{\bf 2}}}

\def\three{{{\bf 3}}}
\def\two{{{\bf 2}}}
\def\one{{{\bf 1}}}

\theoremstyle{definition}

\def\beq{\begin{equation}}
\def\eeq{\end{equation}}
\def\bea{\begin{eqnarray}}
\def\eea{\end{eqnarray}}
\def\bitem{\begin{itemize}}
\def\eitem{\end{itemize}}
\newc{\ie}{{\it i.e.}}          \newc{\etal}{{\it et al.}}
\newc{\eg}{{\it e.g.}}          \newc{\etc}{{\it etc.}}
\newc{\cf}{{\it c.f.}}

 \numberwithin{equation}{section}

\newcommand\fverb{\setbox\fverbbox=\hbox\bgroup\verb}
\newcommand\fverbdo{\egroup\medskip\noindent%
            \fbox{\unhbox\fverbbox}\ }
\newcommand\fverbit{\egroup\item[\fbox{\unhbox\fverbbox}]}
\newbox\fverbbox

\usepackage{authblk}

\begin{document}

\author[1]{Nathaniel Craig\thanks{ncraig@physics.ucsb.edu}}
\author[2,3]{Simon Knapen\thanks{smknapen@lbl.gov}}
\author[4]{Pietro Longhi\thanks{longhi@physics.rutgers.edu}}

\affil[1]{\small{Department of Physics, University of California, Santa Barbara, CA 93106}}
\affil[2]{Berkeley Center for Theoretical Physics,
University of California, Berkeley, CA 94720
}
\affil[3]{ Theoretical Physics Group, Lawrence Berkeley National Laboratory, Berkeley, CA 94720
}
\affil[4]{ New High Energy Theory Center, Rutgers University, Piscataway,
  NJ 08854}

\title{The Orbifold Higgs}

\maketitle

\begin{abstract}
We introduce and systematically study an expansive class of ``orbifold Higgs'' theories in which the weak scale is protected by accidental symmetries arising from the orbifold reduction of continuous symmetries. The protection mechanism eliminates quadratic sensitivity of the Higgs mass to higher scales at one loop (or more) and does not involve any new states charged under the Standard Model. The structures of the Higgs and top sectors are universal and determined exclusively by group theoretical considerations. The twin Higgs model fits within our framework as the simplest example of an orbifold Higgs. Our models admit UV completions as geometric orbifolds in higher dimensions, and fit naturally within frameworks of low scale gauge coupling unification. 

\end{abstract}

\vspace{4cm}

\pagebreak


\vspace{0.2cm}
\noindent

\tableofcontents

\section{Introduction}
With the discovery of a Standard Model-like Higgs boson with a mass near 125 GeV \cite{Aad:2012tfa,Chatrchyan:2012ufa}, experiments are for the first time directly probing the scale at which the electroweak symmetry is broken. At the same time the LHC has imposed impressive constraints on new states near the weak scale, placing increasing stress on conventional paradigms for natural electroweak symmetry breaking.  In light of LHC limits, a crucial question is whether there still remains room for natural mechanisms to stabilize the weak scale. A particularly interesting observation along these lines is that almost all constraints on conventional mechanisms for stabilizing the weak scale (such as composite Higgs models or supersymmetry) hinge critically on the large production cross section for the colored partner particle(s) of the top quark.  The reason for this is simple: If the Higgs is to be protected by a symmetry of some kind, then the top quark must also transform under this symmetry in order for the symmetry to be compatible with a large top yukawa coupling. Popular realizations such as supersymmetry or global symmetries commute with QCD, which ensures that the top's partner particle(s) must carry QCD quantum numbers as well. Given the strong constraints on colored particles from the LHC, it is particularly important to systematically map out the exceptions to this `top partner theorem', since the collider signatures of such exceptions fall in a qualitatively different regime which is still largely unexplored.

There exist several known loopholes in the top partner theorem, all of which hinge on accidental realizations of either a global symmetry (as in twin Higgs models \cite{Chacko:2005pe} and their relatives \cite{Chacko:2005vw, Poland:2008ev, Cai:2008au,Batra:2008jy}), or supersymmetry (as in folded supersymmetry models \cite{Burdman:2006tz}). The question remains, however, whether these loopholes are theoretical curiosities or examples of a very general framework for protecting the weak scale without introducing new states charged under the Standard Model. In this paper we commence the systematic study of loopholes using global symmetries, and show that all the essential features neatly fit within the framework of field-theoretic orbifolds. As a consequence, we identify a large class of models of which the twin Higgs is merely the simplest example. In all of our models, a gauged symmetry group is broken by an orbifold projection, resulting in effective theories including the Standard Model sector plus one or more hidden sectors. Crucially, the effective theory exhibits a continuous \emph{accidental} symmetry relating the Standard Model and hidden sector(s), which suffices to protect the Higgs mass from large quantum corrections. A central feature of this symmetry is that it only involves top partners that are \emph{neutral} under the Standard Model interactions, thus evading the top partner theorem. More generally, {\it all} the important partner partners (such as gauge partners of $SU(2)_L$) are neutral under the Standard Model. We argue that concrete realizations of this idea are strongly constrained by group theoretic considerations, and exploit this observation to initiate a systematic classification of these models.

The most basic example of our class of models, consisting of a $\IZ_{2}$ orbifold, coincides with the well-known twin Higgs model. In this standard scenario \cite{Chacko:2005pe} the physical Higgs is one out of $7$ goldstones arising from the spontaneous breaking of a global $SU(4)$ to $SU(3)$, while the remaining $6$ are eaten by gauging an $SU(2)\times SU(2)$ subgroup of the global $SU(4)$. One of these $SU(2)$ nodes can then be identified with the Standard Model weak gauge group, while the other is its `twin' counterpart. Gauging the $SU(2)\times SU(2)$ subgroup necessarily constitutes an explicit breaking of the $SU(4)$ of which the Higgs is supposed to be a goldstone boson. The Higgs may nevertheless be protected at one loop\footnote{In fact, the protection mechanism in the case of the twin Higgs extends to all loops, but the theory still requires a UV completion in the form of compositeness or supersymmetry around 5-10 TeV.}, if the following criteria are satisfied:
\begin{itemize}
\item 
The $SU(4)$ is a symmetry of the quadratic and quartic parts of the tree-level Higgs potential.
\item The $SU(4)$ is preserved in the quadratic part of the one-loop Higgs potential.
\end{itemize}
In the twin Higgs model, the first criterion is an ad hoc assumption, while the latter is ensured by a discrete $\mathbb{Z}_2$ symmetry which interchanges both $SU(2)$ nodes. Concretely, this discrete symmetry ensures that gauge and yukawa couplings of both groups must be equal, which implies that the leading contributions to the quadratic potential at one loop  take the form
\begin{align}
V(h_A,h_B)&\supset \frac{\Lambda^2}{16\pi^2}\left[ \left( \frac{9}{4} g_A^2 - 6 y_{t,A}^2 \right) |h_A|^2+ \left( \frac{9}{4} g_B^2  - 6 y_{t,B}^2  \right)|h_B|^2\right] \nonumber\\
&= \frac{\Lambda^2}{16\pi^2} \left( \frac{9}{4} g^2 - 6 y_t^2 \right) \left(  |h_A|^2+|h_B|^2\right)\label{quadraticpotential}
\end{align}
where the subscripts `$A$' and `$B$' denote the Standard Model and the twin sector respectively. Equation (\ref{quadraticpotential}) is manifestly $SU(4)$ invariant, and as such the SM-like Higgs (as a pseudo-goldstone of spontaneous $SU(4)$ breaking) is insensitive to the cutoff at one loop.\footnote{Of course, here the uniform cutoff is merely a proxy for physical thresholds that respect the $\mathbb{Z}_2$ symmetry, as is expected of a UV completion such as compositeness or supersymmetry.}

While in the twin Higgs paradigm both the $\mathbb{Z}_2$ and the accidental $SU(4)$ are somewhat artificial ingredients, we stress that both can be \emph{natural} features from the viewpoint of orbifold Higgs models. Specifically, by viewing the twin Higgs as the {orbifold projection}
\begin{equation}
SU(4)/ \mathbb{Z}_2 \rightarrow SU(2)\times SU(2)\times U(1)\,,
\end{equation}
the global $\mathbb{Z}_2$ follows \emph{automatically}. The accidental $SU(4)$ may or may not follow automatically, depending on the spectrum of operators in the ultraviolet.  As we will see, this framework can be naturally extended to include both the top quarks and the Standard Model $SU(3)$ group. Even at the level of the twin Higgs, framing the model in terms of a field theory orbifold is extremely useful. Among other things, it helps to answer the question of how, precisely, the twin Higgs can be distinguished from composite Higgs models at the group theory level: The global symmetry breaking pattern is approximately $SU(4)/SU(3)$, while the field content corresponds to $SU(4)/\mathbb{Z}_2$; this suffices to preserve the $SU(4)$ under modest radiative corrections, while eliminating partner states charged under the Standard Model and introducing a custodial symmetry. It is in this $SU(4)/\mathbb{Z}_2$ that the twin Higgs differs crucially from its composite Higgs cousins.

From a more formal point of view, the utility of orbifolds in controlling large quantum corrections may not be so surprising. In particular, it is well known that in the large-$N$ limit the correlation functions of a daughter theory obtained by orbifolding a mother theory must be identical to the correlation functions of the mother, up to a rescaling of the coupling constants \cite{Bershadsky:1998cb, Kachru:1998ys,Schmaltz}. In the example of the orbifold Higgs, one might expect that the Higgs two-point correlation function must be identical (up to possibly $1/N$-suppressed corrections) to the two-point function in the mother theory, which does enjoy the full protection of the global $SU(4)$. Ultimately we will encounter important subtleties regarding orbifold correspondence in these models, but the conceptual inspiration provided by orbifold correspondence is extremely valuable. Given that the proof provided in \cite{Schmaltz} is valid beyond the simple example of a $\mathbb{Z}_2$ orbifold, the orbifold interpretation of the twin Higgs opens up new possibilities for generalizing beyond this simple case. In fact we will see that the twin Higgs is merely the tip of the iceberg and that a large class of qualitatively new models is waiting to be explored. In this respect, orbifold field theories provide a framework for determining the complete generalization of the twin Higgs mechanism.

In this paper we elaborate on the schematic picture articulated in \cite{Craig:2014aea} and develop the complete framework necessary to construct orbifold Higgs models. We proceed as follows: section \ref{sec:general-features} contains a model-independent analysis of a class of orbifolds by \emph{regular embeddings} of a generic discrete group $\G$, providing a toolkit for the analysis carried out in later sections. We pay particular attention to the breaking of gauge and global symmetries, the rescaling of couplings, and the consequent accidental symmetries of daughter theories. 
In section \ref{sec:examples} we analyze in detail a class of toy models resembling top-Yukawa sectors. These examples feature all essential characteristics of phenomenological interest, and we will easily analyze both abelian and non-abelian orbifolds, building on techniques developed in section \ref{sec:general-features}. In section \ref{sec:lsm} we then study the vacuum of orbifold Higgs models and trace how a Standard Model-like pseudo-goldstone Higgs arises from symmetry breaking in sectors related by orbifold projection. Section \ref{sec:UV-completions} discusses realistic models, touching upon another interesting aspect of field-theoretic orbifolds, namely their amenability to higher-dimensional UV completions as \emph{geometric} orbifolds. Finally, we discuss how phenomenology implies a surprising connection to low scale unification and conclude with discussion of future directions.

\section{Orbifolding General Field Theories }\label{sec:general-features}

A field-theoretic orbifold of a \emph{mother} theory with a discrete symmetry $\G$ consists of projecting onto states invariant under $\G$, a procedure also known as ``gauging $\G$''. At the lagrangian level, the orbifold is realized by retaining field components which are $\G$-invariant. In general this procedure will break both global and gauge symmetries of the mother theory. In this section we review the framework of field-theoretic orbifolds in detail and derive some general results which we will use in later sections, greatly simplifying the tasks of carrying out orbifolds and studying their low energy dynamics. 
The first main result of this section are the general formulae (\ref{eq:symmetry-breaking}), (\ref{eq:surviving-gluons}), (\ref{eq:bifund-invt}) that provide a direct description of the \emph{daughter} theories arising from a given mother theory and discrete symmetry $\G$.

The second main point is a general analysis of the accidental symmetries of the Higgs potential in the daughter theory. 
As we recalled in the Introduction, the twin Higgs model enjoys a realization as a $\IZ_{2}$-orbifold. A considerable advantage of this realization is the fact that the $SU(4)$ accidental symmetry can arise naturally from the $\IZ_{2}$ orbifold. In section \ref{sec:accidentalsym} we will show that this crucial feature extends to field-theoretic orbifolds by a general discrete group $\G$.

As our objective is to establish the structure of an orbifold field theory for general group $\G$, the discussion in this section is necessarily somewhat technical. Readers exclusively interested in the phenomenology of orbifold Higgs models may proceed directly to section \ref{sec:examples}.

\subsection{Background and notation}
Much of the material in this subsection is a review of known results (see for example \cite{Schmaltz}), nevertheless we take the opportunity to set our notation and emphasize a few points which are crucial for our story. 
Given a theory with some gauge or flavor symmetry group $G$, there can be several ways of taking the orbifold, depending on how $\G$ acts on $G$. We will follow \cite{Schmaltz} and stick with the \emph{regular representation} embedding\footnote{The word ``embedding'' is slightly inappropriate, but we will consciously abuse terminology and stick to it. It would be more appropriate to say that we define a $\G$-action on the vector spaces furnishing representations of $G$.}.
Recall that for a finite group $\G=\{g_{1},\dots,g_{\Gamma}\}$ of order $|\G|=\Gamma$,  the regular representation is $\Gamma$ dimensional and simply describes the action of $\G$ onto itself. Concretely, one has a set of $\Gamma\times\Gamma$ matrices acting as permutations on the group elements of $\G$, which are represented by $\Gamma$-dimensional normalized vectors with a single nonzero entry. There are $\Gamma$ such matrices, which we denote by $\gamma^s$, with $s=1,\cdots, \Gamma$. 
The regular representation is reducible and enjoys the well-known decomposition
\begin{equation}\label{eq:regular-decomposition}
\gamma^s=\bigoplus_\alpha \mathbb{1}_{d_\alpha}\otimes r^s_\alpha
\end{equation}
where $\alpha$ runs over all the irreps of $\G$ and $d_\alpha$ is the dimension of the irrep $r_{\alpha}$. This decomposition enjoys the special feature that the multiplicity of each irrep in (\ref{eq:regular-decomposition}) equals its dimension, which will be essential for the rest of our story. Moreover, note that the famous identity $\Gamma = \sum_\alpha (d_\alpha)^2$ is a trivial consequence of  (\ref{eq:regular-decomposition}).

With phenomenological applications in mind, we will restrict ourselves to embeddings of $\G$ in $G=SU(\Gamma N)$ for some positive integer $N$. In particular, let us consider
\be\label{irrepdecom}
\begin{split}
	\gamma^s_{N}& := \mathbb{1}_{N}\otimes \gamma^{s} = \bigoplus_\alpha \mathbb{1}_{N d_\alpha}  \otimes r^s_\alpha\, 
	= \left(\begin{array}{ccc}
	{\ddots} & \Ldash{} & \\
	\cline{1-2}
	& %
	\LRdash{%
	\begin{array}{ccc}
	r_{\alpha}^{s}& & \\
	& \ddots& \\
	& & r_{\alpha}^{s} \\
	\end{array}%
	}%
	& \\
	\cline{2-3}
	& \Rdash{%
	\stackrel{\underbrace{\qquad\qquad\quad}_{N d_{\alpha} \text{ times}}}{\quad}
	} %
	& {\ddots} 
	\end{array}\right)\,.
\end{split}
\ee
Fields  $Q$ and $A$ in the fundamental and adjoint representations of $SU(N\Gamma)$, respectively, transform under $\G$ as 
\begin{equation} \label{GNreps}
	Q\rightarrow \gamma_{N}^s Q\quad\quad A\rightarrow \gamma_{N}^s A (\gamma_{N}^s)^\dagger\,.
\end{equation}
The $\G$-action is obviously {reducible}, and induces the following decomposition of the vector space $V$ of the fundamental representation
\begin{equation}
V=\bigoplus_\alpha V_{\alpha}\otimes R_\alpha
\end{equation}
with 
\begin{equation}
\mathrm{dim}_{\mathbb{C}}(V_\alpha)=N d_\alpha\,,\qquad \mathrm{dim}_{\mathbb{C}}(R_\alpha)= d_\alpha\,,
\end{equation}
where $\G$ acts on the subspace $R_\alpha$ through the irrep $r^s_\alpha$ and leaves $V_\alpha$ invariant. 

In order to capture all the details of orbifolding, it turns out to be useful to introduce a multi-index notation,
according to the patterns of symmetry breaking. Letting $n_{\G}$ be the number of irreps of $\G$, we define the following multi-index $A$, for the vector space $V$:
\be
\begin{split}
	& A \sim (\alpha,a,\a) \qquad \left\{%
	 \begin{array}{l}
		\alpha = 1 \dots n_{\G} \qquad  \text{irrep of $\G$ in the decomposition of $\gamma$} \\
		a = 1\dots d_{\alpha}N \qquad \text{ index for $V_{\alpha}$  }\\
		\a = 1\dots d_{\alpha} \qquad \text{  index for $R_{\alpha}$ }
	 \end{array}
	 \right.\,.
\end{split}
\ee
Note that $\a$ will only be a relevant index if $\G$ is non-abelian. For example, taking $\G=\mathbb{Z}_{\Gamma}$ there are $\Gamma$ one-dimensional irreps, hence the multi-indices span 
\be
	A=(\alpha,a),\qquad \alpha=1,\dots\Gamma,\qquad a=1,\dots N
\ee
with  $\a$ being  suppressed because $d_{\alpha}=1,\ \forall\alpha$.

The discrete group $\G$ may be embedded in more than one gauge group, or both in gauge and flavor groups. For example consider a flavor group $SU(\Gamma F)$ with $F$ a positive integer, and let $W$ be the vector space of the fundamental representation. $\G$ acts reducibly, inducing the decomposition $W= \bigoplus_\mu W_{\mu}\otimes R_\mu $. Correspondingly we introduce multi-indices 
\be
\begin{split}
	& M \sim (\mu,m,\m) \qquad \left\{%
	 \begin{array}{l}
		\mu = 1 \dots n_{\G} \qquad  \text{irrep of $\G$ in the decomposition of $\gamma$} \\
		m = 1\dots d_{\mu}F \qquad \text{ index for $V_{\alpha}$  }\\
		\m = 1\dots d_{\mu} \qquad \text{  index for $R_{\alpha}$ }
	 \end{array}
	 \right.\,.
\end{split}
\ee
Notice in particular that $\mu$ and $\m$ have the same span as $\alpha$ and $\a$ from the orbifolding of the gauge group, while $m$ does not.

\subsection{Projecting onto invariant field configurations}

Having specified how $\G$ acts on the fields in the class of theories of interest, we now turn to describing the $\G$-invariant degrees of freedom singled out by the orbifold. 

Starting with gauge fields, their invariant components under the $\G$-action must obey
\begin{equation}\label{eq:gluon-orbifold}
A= \gamma_{N}^{s}\, A\, (\gamma_{N}^{s})^{\dagger} \quad\Leftrightarrow \quad A\, \gamma_{N}^{s}= \gamma_{N}^{s}\, A 
\end{equation}
The invariant gluons fall into two categories. The first type follows from a direct application of Schur's lemma:
\begin{equation}
A = \bigoplus_\alpha A_{\alpha}\otimes \mathbb{1}_{d_\alpha}\,,
\end{equation}
since $\gamma_{N}^{s}$ has the block-diagonal form exhibited in (\ref{irrepdecom}).
The second type takes the form
\be\label{extraU1}
	{B}_{\vec b} = \bigoplus_{\alpha} \mathbb{1}_{d_{\alpha}N}\otimes \big(b_{\alpha}\cdot\mathbb{1}_{d_{\alpha}}\big)
\ee
where $b_{\alpha}\in\IR$ and $\sum_{\alpha} d_{\alpha}^{2}\cdot b_{\alpha} = 0$ , ensuring that $B_{\vec b}$ is Hermitean and traceless, hence a generator of $SU(\Gamma N)$. Specifically, these are just those elements of the Cartan subalgebra of $SU(\Gamma N)$ that commute with $\gamma_{N}^{s}$.
It is easy to see that there are precisely $n_{\G}-1$ invariant photons of the $B_{\vec{b}}$ type, 
generating $U(1)^{ n_{\G}-1}$. 

Overall, for the regular embedding-type of orbifold, the gauge symmetry breaks to 
\be\label{eq:symmetry-breaking}
	SU(\Gamma N) \quad \longrightarrow \quad \left(\prod_{\alpha=1}^{n_{\G}}SU(d_{\alpha} N) \right)\times \Big( U(1) \Big)^{n_{\G}-1}\,.
\ee
Note how the symmetry breaking pattern is fully determined by group-theoretical properties of the orbifold group: each non-abelian factor corresponds to an irreducible representation of $\G$, with the dimensionality $d_\alpha$ determining the corresponding rank. If $\mathcal{G}$ is abelian, and therefore $d_\alpha=1\; \forall \alpha$, then the rank of the daughter symmetry group as a whole equals the rank of the mother symmetry group. For non-abelian $\G$, the rank of the daughter is always less than the rank of the mother.
In what follows we will mainly focus on the non-abelian factors in (\ref{eq:symmetry-breaking}), while neglecting the abelian ones; the physical motivations behind this choice are discussed in section \ref{sec:UV-completions}.
For later convenience we also express the surviving gluons in multi-index notation, as degrees of freedom of the mother theory
\be\label{eq:surviving-gluons}
\begin{split}
	&A_{A}^{\phantom{A}\,B} 
	= (A^{(\alpha)})_{a}^{\phantom{a}\, b}\,\delta_{\alpha}^{\phantom{\alpha}\beta}\,\delta_{\a}^{\phantom{\a}\b} \,.
\end{split}
\ee

Turning to matter fields in a generic representation $R$ of $G$, the projector onto the invariant subspace reads
\begin{equation}
P_R=\frac{1}{\Gamma}\sum_{s=1}^\Gamma \gamma^s_R
\end{equation}
where $\gamma_{R}^{s}$ is the matrix representation of $g_{s}$.
With an eye towards subsequent applications, we will focus on a field in the bifundamental of  $SU(\Gamma N)\times SU(\Gamma F)$. With the regular embedding described above, the $\G$-action on such a field will be expressed by matrices%
\footnote{For clarity on our conventions, in vector-matrix notation we take an anti-fundamental tensor to transform as $\psi \mapsto \gamma^{*}\cdot \psi$, such that contraction with a fundamental $\lambda$ yields an $SU(N)$ invariant $\psi^{T}\cdot \lambda \mapsto \psi^{T}\cdot(\gamma^{*})^{T}\cdot \gamma\cdot\lambda$. In our index notation we write $\psi^{A}\mapsto\psi^{B} (\gamma^{\dagger})_{B}^{\phantom{B}A} = (\gamma^{*})^{A}_{\phantom{A}B}\psi^{B}$. This explains the appearance of complex conjugation in (\ref{eq:reg-bifund}): in building a left-acting projection operator we think of both groups $SU(\Gamma N)\times SU(\Gamma F)$ as acting from the left.}
\be\label{eq:reg-bifund}
\gamma_{N\otimes F}^s = \gamma_N^s\otimes \left(\gamma_F^s\right)^{*}
\ee
with $\gamma_N^s$ and $\gamma_F^s$ the regular embeddings of $\G$ in the the two symmetry groups.

\begin{figure}[h!]
\begin{center}
\includegraphics[width = 0.9\textwidth]{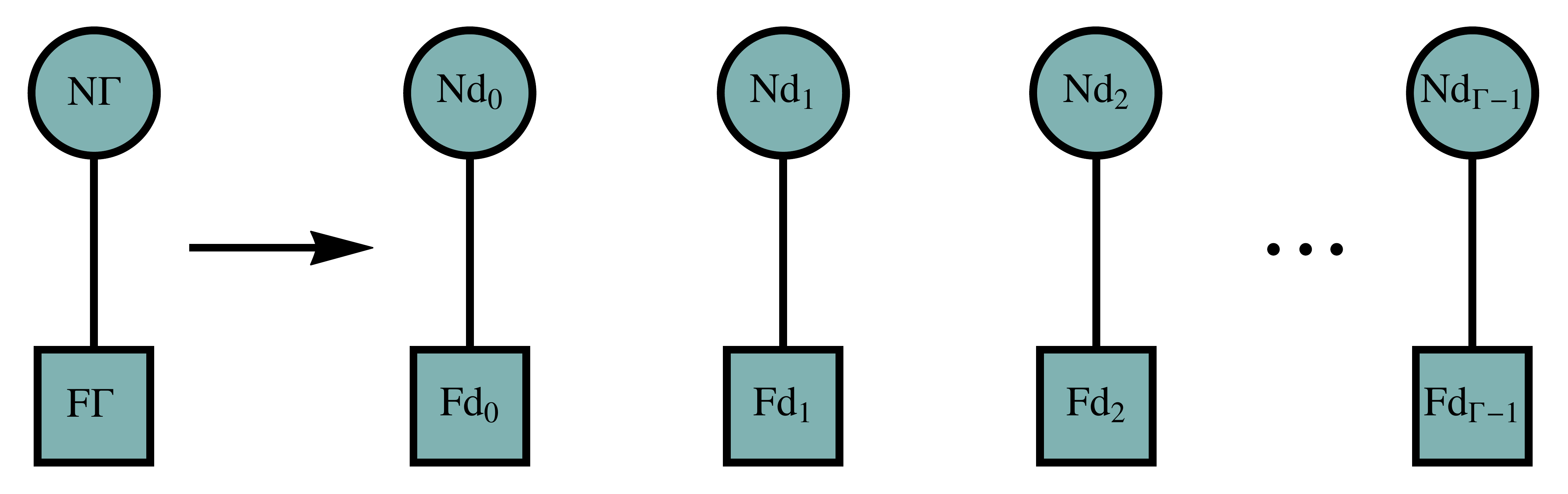}
\caption{The orbifold breaking of a mother theory with a single bi-fundamental field into daughter nodes. Here circles and squares represent gauge and global symmetries, respectively. }
\label{fig:simplequiver}
\end{center}
\end{figure}

According to the orthogonality theorem for matrix elements of irreducible representations, we may write the bifundamental projector as
\be
\begin{split}\label{projectorOpp}
	\big(P_{N\otimes F}\big)_{A\phantom{AB,}N}^{\phantom{M}MB} %
	& = \delta_{\alpha}^{\phantom{\alpha}\beta}(\mathbb{1}_{ N d_\alpha})_{a}^{\phantom{\a}b}  \cdot \delta^{\mu}_{\phantom{\alpha}\nu}(\mathbb{1}_{F d_\mu})^{m}_{\phantom{\a}n} \cdot \left( \frac{1}{\Gamma}\sum_s (r^{s}_{\alpha})_{\a}^{\phantom{\a}\b}\cdot (r^{s\, *}_{\mu})^{\m}_{\phantom{\a}\n} \right) \\
	& = \frac{1}{{d_{\alpha}}}\,\delta_{\alpha}^{\phantom{\alpha}\beta}\delta^{\mu}_{\phantom{\alpha}\nu}\,\delta_{a}^{\phantom{\a}b}  \delta^{m}_{\phantom{\a}n} \,\delta_{\alpha}^{\mu} \delta_{\a}^{\m}\delta^{\b}_{\n}
\end{split}
\ee
where in the first line we expressed the Kronecker-$\delta$ symbols as unit matrices to highlight the fact that the span of the corresponding indices is actually $\alpha$- (resp. $\mu$-) dependent. Although the notation of the second line is somewhat abusive, we will often make use of it, keeping this remark in mind.
From (\ref{projectorOpp}) we can write down in full generality the invariant components of a bifundamental tensor $\Phi_{A}^{\phantom{A}M}$
\be\label{eq:bifund-invt}
\begin{split}
	(\phi^{(\alpha)})_{(a,\a)}^{\phantom{(a,\a)}(m,\m)} \,:=\, \big(P_{N\otimes F}\big)_{A\phantom{AB,}N}^{\phantom{M}MB} \, \Phi_{B}^{\phantom{A}N} & = \frac{1}{d_{\alpha}}\, \delta_{\alpha}^{\mu}\,\delta_{\a}^{\m}\,\sum_{\a'=1}^{d_{\alpha}}\Phi_{(\alpha,a,\a')}^{\phantom{(\alpha,m,\a')}(\alpha,m,\a')} \,,
\end{split}
\ee
where the indices span
\be\label{eq:indexspan}
	\alpha,\mu=1,\dots,n_{\G}\quad \a,\m=1,\dots,d_{\alpha}\quad a=1,\dots,Nd_{\alpha}\quad m=1,\dots,Fd_{\alpha}\,.
\ee
This is where the role of multi-indices as a valuable book-keeping device becomes manifest: in expression (\ref{eq:bifund-invt}) we can read off \emph{directly} how invariant field components transform under the symmetries of the daughter theory. 
In particular, besides the breaking of the gauge symmetry described above in (\ref{eq:symmetry-breaking}), we now find a similar breaking pattern for the flavor symmetry as well
\be\label{eq:flavor-breaking}
	SU(\Gamma F) \quad \longrightarrow \quad \left(\prod_{\alpha=1}^{n_{\G}}SU(d_{\alpha} F) \right)\times \Big( U(1) \Big)^{n_{\G}-1}\,.
\ee
This simple example is conveniently summarized in figure \ref{fig:simplequiver}. 

\subsection{Scaling of couplings}\label{sec:scaling} 
The orbifold establishes a precise relation between the couplings of the mother theory and those of the daughter theory. In this section we analyze the scaling of various types of couplings of interest to us. We will mainly be interested in applications involving a gauge theory with matter fields transforming as bi-fundamentals of two symmetry groups in which $\G$ is embedded through the regular representation (or in slightly different setups which happen to share the same conclusions).

From the general formulae (\ref{eq:surviving-gluons}) and (\ref{eq:bifund-invt}) it is clear that each mother field will descend to a daughter field for each irrep of $\G$, labeled by $\alpha$.  Kinetic terms of the mother theory's lagrangian generally descend to \emph{unnormalized} kinetic terms for daughter fields, inducing a canonical rescaling of the field strengths; this is the underlying mechanism driving a corresponding rescaling of all couplings.

In particular, we can read off the scaling factors introduced by the projection:
\begin{itemize}
\item every bi-fundamental comes with a coefficient $1/d_\alpha$
\item gauge fields do not come with any multiplicative factor
\item each trace contributes%
\footnote{The presence of $\delta_{\a}^{\m}$ in (\ref{eq:bifund-invt}) is responsible for this factor.}
a coefficient $d_\alpha$.
\end{itemize}
With these in mind, it is easy to see how various operators in the mother theory project down to the daughter:

\be
\begin{split}
\mathrm{ Tr}\,\partial \Phi^{\dagger}\partial \Phi&\rightarrow \frac{1}{d_\alpha} \,\mathrm{Tr}\, \partial \phi^{(\alpha)\,\dagger}\partial \phi^{(\alpha)}\\
m\, \mathrm{ Tr}\, \Phi^{\dagger}\Phi&\rightarrow \frac{m}{d_\alpha}\, \mathrm{ Tr}\, \phi^{(\alpha)\,\dagger}\phi^{(\alpha)}\\
\qquad \qquad %
\lambda\, \big( \mathrm{ Tr}\,\Phi^{\dagger}\Phi\big)^{2}&\rightarrow %
\lambda \, \left( \frac{1}{d_\alpha}\mathrm{ Tr}\,  \phi^{(\alpha)\,\dagger} \phi^{(\alpha)}\right)^{2} \\%
\delta\,\mathrm{ Tr}\, \Phi^{\dagger}\Phi\Phi^{\dagger}\Phi &\rightarrow \frac{\delta}{d_\alpha^3} \,\mathrm{ Tr}\,\phi^{(\alpha)\,\dagger}\phi^{(\alpha)}\phi^{(\alpha)\,\dagger}\phi^{(\alpha)}\\
y\,\mathrm{ Tr}\,  \Phi\Psi\Psi' &\rightarrow  \frac{y}{d_\alpha^2}\,\mathrm{ Tr}\,  \phi^{(\alpha)} \psi^{(\alpha)}{\psi'}^{(\alpha)} \\
\frac{1}{g^2}\,\mathrm{Tr}FF&\rightarrow\frac{d_\alpha}{g^2}\,\mathrm{Tr}F^{(\alpha)}F^{(\alpha)}.
\end{split}
\ee
where $\Phi, \Psi$ and $\Psi'$ are bi-fundamental fields of the mother theory, while $\phi_\alpha, \psi_\alpha$ and $\psi_\alpha'$ are the corresponding daughter fields attached to the $\alpha$-th node. Sums over repeated greek indices are understood.

Finally, upon restoring the canonical normalization of the kinetic terms, the various coupling constants in the daughter theory must therefore be rescaled as follows:
\be
\begin{split}\label{eq:rescalecoupling}
\phantom{\lambda}m\rightarrow m\phantom{\lambda} \quad\quad y &\rightarrow  \frac{y}{\sqrt{d_\alpha}}\\
\phantom{m}\lambda\rightarrow\lambda\phantom{m}\quad\quad \delta \to {\delta\over d_{\alpha}} & \quad \quad g \rightarrow  \frac{g}{\sqrt{d_\alpha}}\,.
\end{split}
\ee
The rescaling of the gauge and yukawa couplings may therefore be different for different sectors in the daughter theory, a feature that will play a crucial role in the realization of the accidental symmetry we are after. This is the subject of the next section.

\subsection{Additional symmetries in orbifold models}\label{sec:accidentalsym}
Now that we have seen how various couplings rescale, we are in a position to try and say something general about the symmetries of a daughter theory. 

\subsubsection{Exact and discrete\label{sec:discrete}}
Our daughter theories can typically be described by a quiver structure (such as the one in figure \ref{fig:simplequiver})  involving gauge nodes with possibly different ranks, as dictated by the dimensionality of $\G$-irreps. This pattern extends to global symmetries as well. As a consequence we can say that generally the daughter theory will enjoy a \emph{new kind} of discrete symmetry $\CS_{\vec d}$, which is simply the symmetry group of the tuple of positive integers
\be
	\vec d = (d_{1},d_{2},\dots,d_{n_{\G}})\,.
\ee
For example, in the case that $\G=\IZ_{\Gamma}$, all $d_{\alpha}$ are $1$ and the symmetry group of the quiver is simply $\CS = S_{\Gamma}$. Another example is $\G=A_{4}$ for which there are three irreps of dimension $1$, and one irrep of dimension $3$, hence resulting in a quiver with an $S_{3}$ symmetry group permuting the three ``sectors'' corresponding to $d_{\alpha}=1$.

The origin of $\CS_{\vec d}$ should be clear: an $SU(\Gamma N)$ symmetry of the mother theory has a Weyl subgroup $S_{\Gamma N}$, which gets broken {by the orbifold} to 
\be
	\left(\prod_{\alpha}S_{d_{\alpha}N}\right)\, \times \CS_{\vec d}
\ee
with each factor in the product being a Weyl subgroup of $SU(d_{\alpha}N)$, and the $\CS_{\vec d}$ acting by permutations of different sectors (with $d_{\alpha}=d_{\beta}$). 
{Each symmetry of the mother thus contains such an $\CS_{\vec d}$ subgroup. However, the daughter theory only has a \emph{single} $\CS_{\vec d}$ acting ``diagonally'', i.e. permuting all nodes and bifundamentals at once (see figure \ref{fig:simplequiver}).}
 The explanation comes simply from the projection of bi-fundamental fields: the $\delta^{\mu}_{\alpha}$ in (\ref{eq:bifund-invt}) is what singles out the diagonal 
\be
	\CS_{\vec d}\ \subset\ \CS^{(N)}_{\vec d} \times \CS^{(F)}_{\vec d}\ \subset\ SU(\Gamma N) \times SU(\Gamma F)\,.
\ee
The effects of $\CS_{\vec d}$ are also clear: labeling sectors of the daughter theory by $\alpha=1,\dots,n_{\G}$, for any two sectors $\alpha,\beta$ with $d_{\alpha}=d_{\beta}$, the quantum action will be invariant under switching the labels $(\alpha)\leftrightarrow(\beta)$ on all involved daughter fields. To give an example, any two fields $\phi^{(\alpha)},\phi^{(\beta)}$ descending from the same mother field $\Phi$ are guaranteed to have equal masses to {all orders in perturbation theory} 
\be
	m_{\phi^{(\alpha)}}=m_{\phi^{(\beta)}}\,.
\ee 
In the special case where $\mathcal{G}=\mathbb{Z}_2$, the $\CS_{\vec d}$ is precisely the $\mathbb{Z}_2$ symmetry that protects the Higgs mass in the twin Higgs model.

\subsubsection{Accidental and continuous\label{sec:accidentalsym}}
While for certain orbifolds (such as $\G=\IZ_{\Gamma}$) the manifest $\CS_{\vec d}$ constraint is a powerful one, more general discrete groups will feature irreps of various {different} dimensions, curtailing the effectiveness of $\CS_{\vec d}$.
Surprisingly, another less manifest, but farther reaching feature of orbifolds plays an analogous role.
In addition to the exact discrete symmetry, orbifold models -- with a caveat to be stated presently -- also enjoy an accidental \emph{continuous symmetry}.

More precisely, in a generic mother theory with bi-fundamental scalars $\Phi_{A}^{\phantom{A}M}$  of $SU(\Gamma N)\times SU(\Gamma F)$, the scalar potential will feature the following operators
\be\label{eq:mother-operators}
	{\Phi^{\dagger}}{}_{M}^{\phantom{A}A}\,  \Phi_{A}^{\phantom{A}M}\,,\qquad %
	\big( {\Phi^{\dagger}}{}_{M}^{\phantom{A}A}\, \Phi_{A}^{\phantom{A}M} \big)^{2}\,, %
	\qquad {\Phi^{\dagger}}{}_{M}^{\phantom{A}A}\, \Phi_{A}^{\phantom{A}N} {\Phi^{\dagger}}{}_{N}^{\phantom{A}B}\, \Phi_{B}^{\phantom{A}M} \,,
\ee
whose orbifold projections readily follow from (\ref{eq:bifund-invt}):
\begin{eqnarray}\label{eq:newinvariant}
	{\Phi^{\dagger}}{}_{M}^{\phantom{A}A}\, \Phi_{A}^{\phantom{A}M} %
	&  \ \ \mapsto\ \  &  %
	\sum_{\alpha}({\phi}^{(\alpha)\,\dagger}){}^{\phantom{a} a}_{m}\,(\phi^{(\alpha)})_{a}^{\phantom{a}m} \\
	\label{eq:quartic-single-trace-projected}
	{\Phi^{\dagger}}{}_{M}^{\phantom{A}A}\, \Phi_{A}^{\phantom{A}N} {\Phi^{\dagger}}{}_{N}^{\phantom{A}B}\, \Phi_{B}^{\phantom{A}M} %
	& \ \ \mapsto\ \  &  %
	\sum_{\alpha}\,{1\over d_{\alpha}}({\phi}^{(\alpha)\,\dagger}){}^{\phantom{a} a}_{m}\,(\phi^{(\alpha)})_{a}^{\phantom{a}n} ({\phi}^{(\alpha)\,\dagger}){}^{\phantom{a} b}_{n}\,(\phi^{(\alpha)})_{b}^{\phantom{a}m}\,
\end{eqnarray}
where daughter fields have been canonically rescaled.
Now the RHS of (\ref{eq:newinvariant}) can be seen to enjoy a large, continuous symmetry. Note from (\ref{eq:indexspan}) that the field $\phi^{(\alpha)}$ has $(d_\alpha N)\cdot (d_{\alpha} F)$ independent components, then recalling the identity $\sum_\alpha d_\alpha^2=\Gamma$, we see that the RHS of (\ref{eq:newinvariant}) 
is in fact%
\footnote{An alternative viewpoint is the following. 
The LHS of (\ref{eq:newinvariant}) is an invariant of $SU(\Gamma^{2} N F)$, the effect of the orbifold is to reduce the number of $\Phi$-components by a factor of $\Gamma$, naturally breaking the symmetry to $SU(\Gamma^{} N F)$.} %
an invariant of
\be\label{eqn:accidentalsym}
	\CS_{acc} = SU( \Gamma N F)\,.
\ee
On the other hand (\ref{eq:quartic-single-trace-projected}) breaks $\CS_{acc}$. Hence the caveat is that $\CS_{acc}$ is only realized in those theories for which the single-trace quartic operator is either absent entirely or much smaller than the double-trace quartic.

\begin{figure}
\begin{floatrow}
\ffigbox{%
\includegraphics[width=0.5\textwidth]{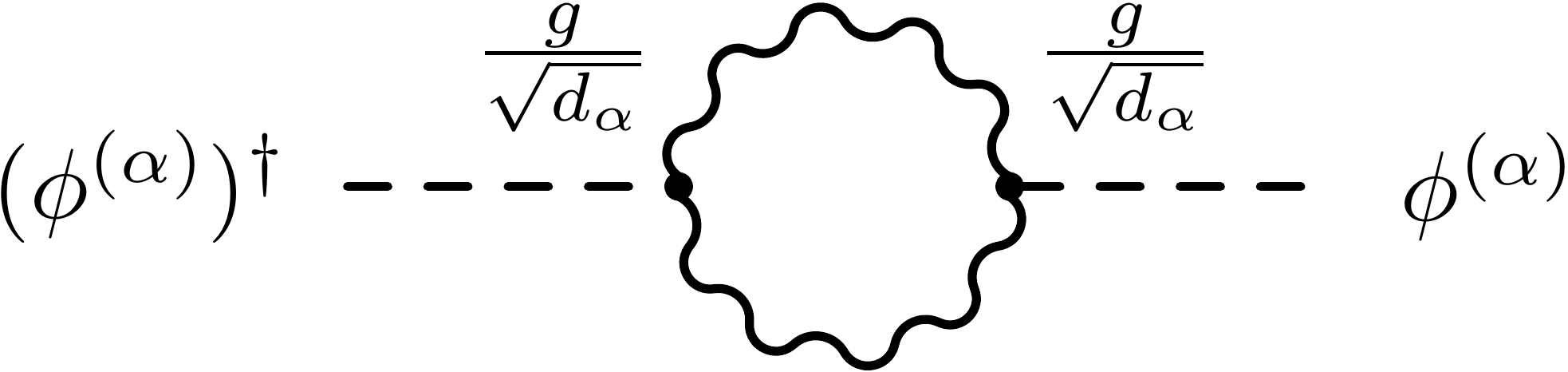} 
}{%
\caption{Example correction the $|\phi^{(\alpha)}|^2$ operator in the daughter theory. The rescaling of the gauge couplings compensates for the $d_\alpha$ dependence in the Casimir.\label{fig:quadratic}}%
}
\hspace{7pt}
\ffigbox{%
\includegraphics[width=0.4\textwidth]{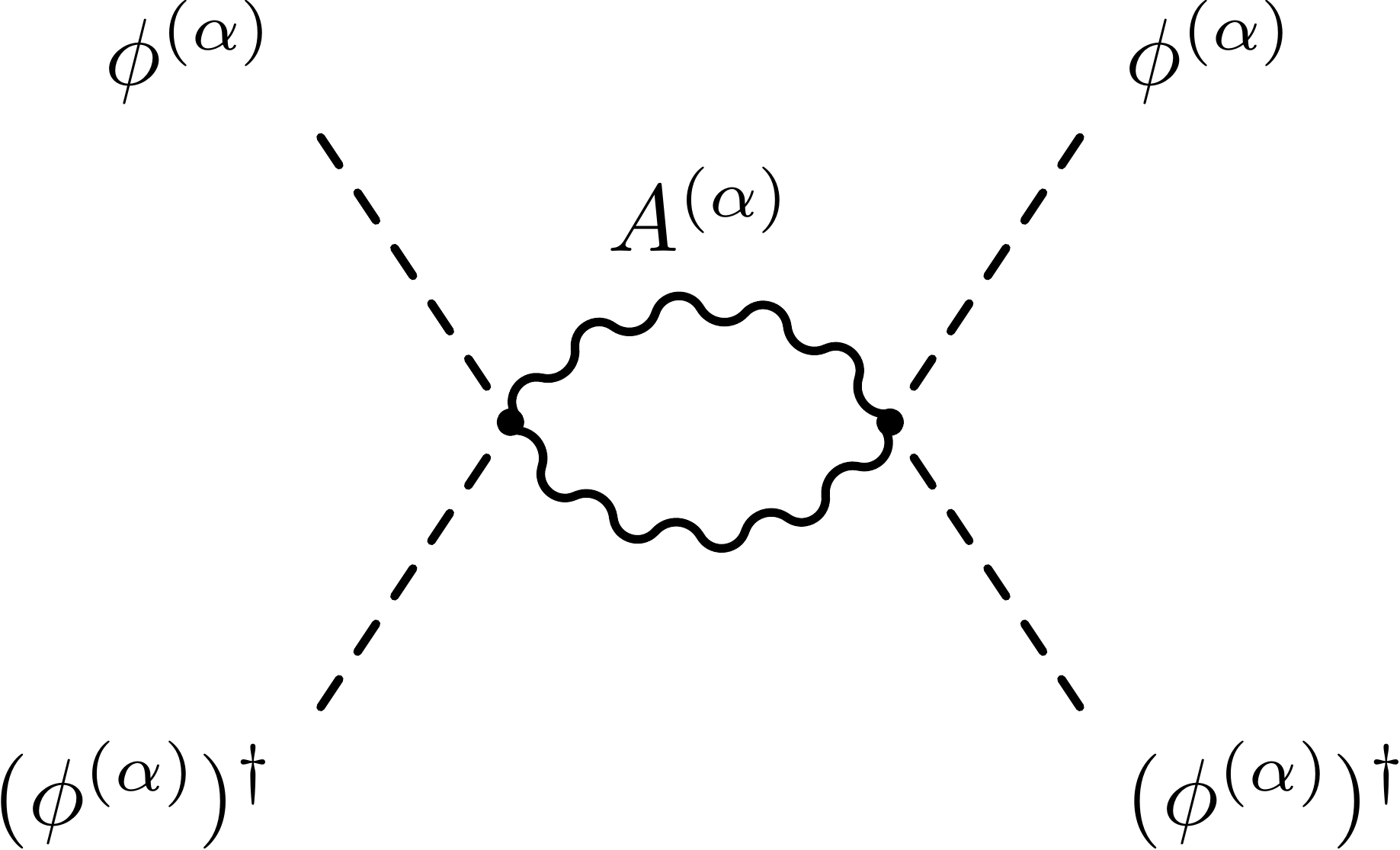} 
}{%
\caption{Example correction to the $|\phi^{(\alpha)}|^4$ in the daughter theory. The diagram only exists if the $\phi^{(\alpha)}$ on all ends belong to the same sector. \label{fig:quartic}}%
}
\end{floatrow}
\end{figure}

We refer to $\CS_{acc}$ as an \emph{accidental} symmetry, since it holds at tree level, and is neither preserved by possible gauge interactions nor by possible single-trace quartics of (\ref{eq:quartic-single-trace-projected})\footnote{This is distinct from ``accidental symmetries'' in the context of the Standard Model, which are ``symmetries'' that arise only because the genuine symmetries of the theory forbid relevant or marginal operators violating the accidental symmetries.}. We mention \emph{en passant} that quite generally for the models we will consider, $\CS_{acc}$ is a symmetry of the whole quadratic action (not just for the scalars), although this fact will not play a role in the rest of this paper.
We now come to the most important feature of $\CS_{acc}$: while it is natural to expect that it be broken by one-loop corrections of the form shown in figure \ref{fig:quadratic}, surprisingly this is not the case for the two-point function!
In fact, with gauge interactions turned on, one would expect that $\phi^{(\alpha)\,\dagger}\,\phi^{(\alpha)}$ and $\phi^{(\beta)\,\dagger}\,\phi^{(\beta)}$ would get corrected differently, in particular when $d_{\alpha}\neq d_{\beta}$. However at this point something special happens: \emph{The $d_\alpha$ dependence from the Casimir cancels against the $d_\alpha$ dependence from the rescaling of the gauge coupling in (\ref{eq:rescalecoupling}), up to a $\frac{1}{N}$ suppressed correction.} Concretely, the contribution to the one-loop effective potential from gauge loops is given by
\be\begin{split}\label{eq:effaction}
V_{eff} &\supset  \frac{3}{16\pi^2}\sum_{\alpha=1}^{n_{\G}}\frac{( d_\alpha N)^2-1}{2 d_\alpha N}\left(\frac{g}{\sqrt{d_\alpha}}\right)^2|\phi^{(\alpha)}|^2\Lambda^2\\
&=\frac{3}{32 \pi^2} g^2   N \left( \sum_{\alpha=1}^{n_{\G}}|\phi^{(\alpha)}|^2 \right) \Lambda^2
- \frac{3}{32 \pi^2} \frac{g^2}{N} \left(\sum_{\alpha=1}^{n_{\G}}\frac{1}{d_\alpha^2}|\phi^{(\alpha)}|^2\right)\Lambda^2\,.
\end{split}
\ee

The first term in (\ref{eq:effaction}) is again manifestly $SU( \Gamma N F)$ symmetric, while from the second term we find a parametrically small breaking of $\CS_{acc}$, of the order
\be
	\delta m^{2} \sim \ \frac{3 g^2 \Lambda^2}{32 \pi^2} \, \frac{1}{N} \left(1 - \frac{1}{d_{n_{\G}}^2}\right)
\ee
where $d_{n_{\G}}$ is understood to be the irrep of $\G$ of largest dimension, and we used the fact that any group $\G$ always admits an irrep of dimension one. Amusingly, in the $\G=\IZ_{\Gamma}$ case (which includes the Twin Higgs), this term vanishes exactly, affording an even softer breaking of $\CS_{acc}$. 

The roots of this crucial result reach deep into the nature of field-theoretic orbifolds: a general theorem asserts that in the large-$N$ limit the correlation functions of a daughter theory should be identical to those of the mother, up to the usual rescaling of the coupling constants \cite{Bershadsky:1998cb, Kachru:1998ys,Schmaltz}. 
In particular, the different mass terms $\phi^{(\alpha)\,\dagger}\phi^{(\alpha)}$ of daughter fields should all be expected to converge to the mother's $\Phi^\dagger \Phi$ operator in the large $N$ limit, compatibly with (\ref{eq:effaction}).
For the sake of clarity, it should be noted that such correspondence does not necessarily entail a full-fledged duality of the two theories\footnote{We thank Aleksey Cherman for bringing this to our attention.}. More precisely, the necessary and sufficient conditions for an actual orbifold equivalence were worked out in \cite{Kovtun:2004bz}, and may be violated in some of our models. This nevertheless does not affect our conclusions, which revolve around the dynamics of the daughter theories, rather than on their putative equivalence to mother theories.

For the quartic potential the situation is different: although it is $\CS_{acc}$-invariant at tree-level, the symmetry is radiatively spoiled for this operator. This can be easily seen: noting that
\be
\begin{split}
(\Phi^\dagger \Phi)^2 \mapsto  & \Big( \sum_{\alpha}  \phi^{(\alpha)\dagger}\phi^{(\alpha)} \Big)^{2}  \\
= & \sum_{\alpha}  (\phi^{(\alpha)\dagger}\phi^{(\alpha)})^2 + \sum_{\alpha\neq\beta}  (\phi^{(\alpha)\dagger}\phi^{(\alpha)})(\phi^{(\beta)\dagger}\phi^{(\beta)})\,,
\end{split}
\ee
at one loop the first sum in the second line gets corrections from gauge interactions as illustrated in figure \ref{fig:quartic}, while no such corrections occur for the second sum. 
This undemocratic treatment of terms is what spoils the symmetry. 
This is not in contradiction with the above statement about the large $N$ limit of correlation functions, since the orbifold correspondence derived in \cite{Schmaltz} only applies to single trace operators. 

\subsection{An orbifold toolkit}
Our analysis of the relevant aspects of field-theoretic orbifolds is now complete. Before turning to applications, let us recollect them in the form of a ``toolkit'', of which we will make extensive use later. 
The recipe goes as follows:
\begin{itemize}
	\item Identify the irreps of $\G$, hence the canonical decomposition of the regular representation (cf. (\ref{eq:regular-decomposition})).
	\item Write down the daughter theory: typically it will consist of a quiver gauge theory with $n_{\G}$ disjoint nodes, each one carrying a structure similar to that of the mother theory. 
	Specifically
	\begin{itemize}
		\item gauge and global symmetries break according to (\ref{eq:symmetry-breaking}),
		\item gauge fields are given by (\ref{eq:surviving-gluons}) and (\ref{extraU1}); matter fields by (\ref{eq:bifund-invt}).
	\end{itemize}
	\item Identify the symmetries of the daughter theory: these include both
	\begin{itemize}
		\item the exact discrete symmetry 
		$\CS_{\vec d}$
		\item the accidental continuous symmetry $\CS_{acc}$: with the caveat about the single-trace quartic stated above, for a mother scalar field in the bifundamental of $SU(\Gamma N) \times SU(\Gamma F)$ one has $\CS_{acc} = SU(\Gamma N F)$. 
	\end{itemize}
\end{itemize}

\section{The Orbifold Higgs}\label{sec:examples}
With all our tools sharpened, we are now ready to apply them to the orbifold Higgs. 
All results will be obtained as straightforward applications of the machinery from section \ref{sec:general-features}, without the necessity to perform any further computations.
In this section we focus on a detailed analysis of orbifold Higgs toy models, consisting only of the gauge, top and Higgs sectors. 
A discussion of additional features of the Standard Model which aren't directly relevant to naturalness will be postponed to section \ref{sec:UV-completions}. 
Consequently, readers should not be too distressed by the anomalous nature of the `model fragments' discussed here, an evil which will be easily cured by putting the rest of the Standard Model back in; we are merely focusing on the Orbifold Higgs equivalent of ``natural supersymmetry''.

\subsection{Setup and discussion}
\begin{figure}
\begin{floatrow}
\ffigbox{%
\includegraphics[width=0.2\textwidth]{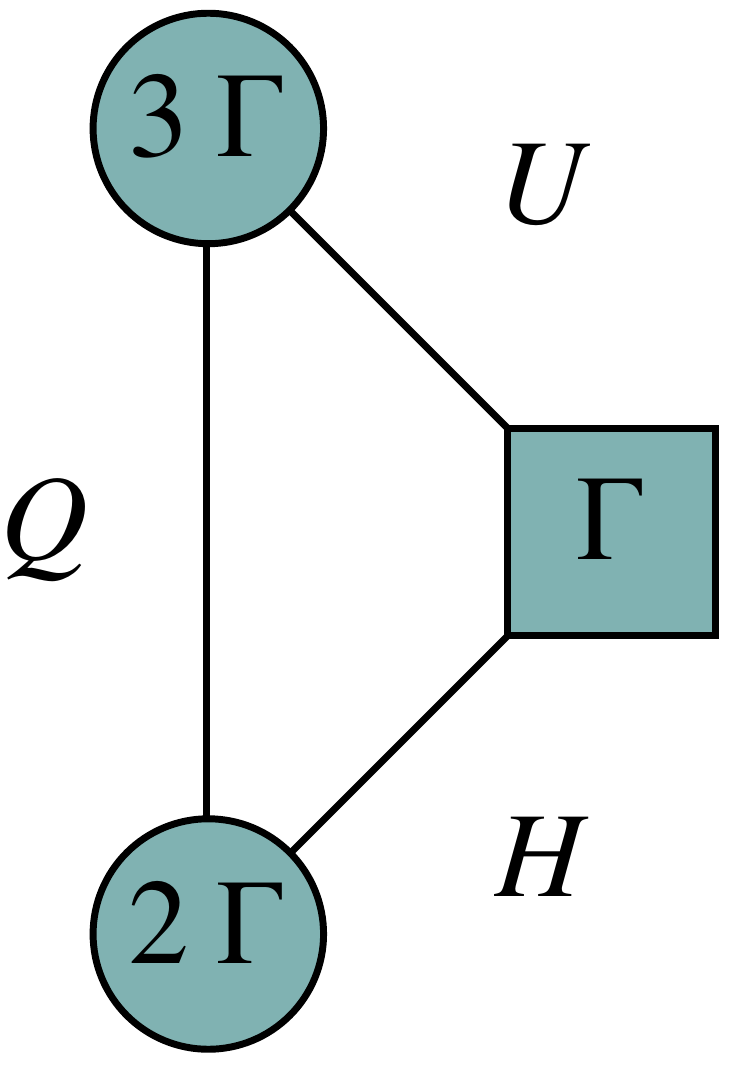} 
}{%
\caption{The quiver diagram of the mother theory. The circles are gauge symmetries, the square represents the flavor symmetry.\label{fig:motherquiver}}%
}
\hspace{15pt}
\capbtabbox{%
\begin{tabular}{c|cc|c}
&$SU(3\Gamma)$ &$SU(2\Gamma)$&$SU(\Gamma)$\\\hline
$H$&$1$&$ \square$& $\overline{\square}$\\
$Q$&$ \square$&$ \overline{\square}$&$1$\\
$U$& $\overline{\square}$&$1$&$\square$
\end{tabular}
}{%
  \caption{Matter content of the mother theory.\label{tab:mothertheory}}%
}
\end{floatrow}
\end{figure}

The general mother theory of Orbifold Higgs models has gauge symmetry $SU(3\Gamma)\times SU(2\Gamma)$ as well as an $SU(\Gamma)$ flavor symmetry (the purpose of the latter will be clarified below in section \ref{sec:bifundamentals}). 
The matter content consists of a scalar $H$ and two fermions $Q$ and $U$, all of which are taken to be bifundamentals in such a way that a gauge-invariant Yukawa interaction term $H Q U$ is allowed.  The content of this theory is summarized by table \ref{tab:mothertheory} and the corresponding quiver diagram in figure \ref{fig:motherquiver}. 
For clarity, we will use multi-indices $A,B$ for $SU(2\Gamma)$, $C,D$ for $SU(3\Gamma)$, and $M,N$ for $SU(\Gamma)$ throughout the rest of this section.

With the discussion of section \ref{sec:accidentalsym} in mind, we take the single trace quartic operator in (\ref{eq:mother-operators}) to be absent in the mother theory\footnote{Note that a vanishing single trace quartic is strictly speaking an unnecessarily strong assumption: For the orbifold Higgs mechanism to function, it in practice suffices that the single trace quartic enjoys a roughly 20\% suppression with respect to the double trace quartic. We will further elaborate on this in section \ref{sec:lsm}.}, and consider the following tree level potential 
\be\label{eq:mother-scalar-potential}
	V[\Phi] \, =\, -m^2\,  {H^{\dagger}}{}_{M}^{\phantom{A}A}\, H_{A}^{\phantom{A}M} \, %
	+ \, \lambda \,\big( {H^{\dagger}}{}_{M}^{\phantom{A}A}\, H_{A}^{\phantom{A}M} \big)^{2}\, + y\; H_{A}^{\phantom{A}M} Q_{C}^{\phantom{B}A}U_{M}^{\phantom{B}C}\,,
\ee
up to irrelevant operators. 
Considering then an orbifold projection by a generic $\G$, it follows directly from the analysis of section \ref{sec:accidentalsym} that the Higgs sector of the daughter theory enjoys the accidental symmetry (\ref{eqn:accidentalsym}). 
In the case at hand  $\CS_{acc} = SU(2\Gamma)$ (where $\Gamma = |\G|$ as usual), 
a further important feature -- on which we will further elaborate below -- is that $\CS_{acc}$ is preserved at one loop in the quadratic action.
We then come to the essence of the Orbifold Higgs: \emph{the daughter theory naturally features all the crucial ingredients at work in the Twin Higgs mechanism, with the immediate generalization that the physical Higgs will be a pseudo-goldstone boson of the accidental $SU(2\Gamma)$}. 
Orbifold Higgs models do in fact provide a neat classification of generalizations of the Twin Higgs: for every discrete group $\mathcal{G}$ one can write down an orbifold Higgs model. The Twin Higgs fits within this classification as the simplest example, with  $\G = \IZ_2$.

In the remainder of this section we will illustrate the mechanism with two explicit examples involving an abelian and a non-abelian orbifold group, respectively. We further comment on the importance of the bifundamentals in section \ref{sec:bifundamentals} and briefly recollect the general results in section \ref{sec:recipe}.

\subsection{Abelian Example}\label{sec:abelian-toy}
The simplest generalizations of the Twin Higgs involve $\G = \IZ_{\Gamma}$ orbifolds. For $\mathbb{Z}_\Gamma$ all $d_\alpha=1$, and (\ref{eq:symmetry-breaking}), (\ref{eq:bifund-invt})  immediately indicate that the daughter theory will simply involve $\Gamma$ copies of the Standard Model Higgs sector. 

Consider the standard multiplicative realization of $\IZ_{\Gamma}$:
\be
	\IZ_{\Gamma} = \{1,\zeta,\dots,\zeta^{\Gamma-1}\}\,,\qquad \zeta=e^{2\pi i/\Gamma}\,,
\ee
where every $\IZ_{\Gamma}$-irrep has dimension $1$ and  they read $r_\alpha^s=\zeta^{s\,\alpha}$
with $s,\alpha=0,\dots\Gamma-1$. The regular representation then takes the explicit form
\be \label{eq:Zgammarep}
	\gamma^{1} = %
	\left(\begin{array}{cccc}
	1 & & & \\
	 & \zeta & & \\
	 & & \ddots & \\
	 & & & \zeta^{\Gamma-1}
	\end{array}\right)\,,%
	\qquad  \gamma_{}^{s} = \big(\gamma_{}^{1}\big)^{s}\,,\qquad s=0,\dots, \Gamma-1 \,.
\ee
For clarity, we also give the explicit form of the regular embedding of $\G$ into $SU(2\Gamma)$ (an analogous formula holding for $SU(3\Gamma)$):
\be
	\gamma_{N=2}^{s} = \left(\begin{array}{cccccc}
	\begin{array}{cc}
	1 & \\
	& 1
	\end{array}%
	& \Ldash{} & & & &\\
	\cline{1-2}
	& %
	\LRdash{%
	\begin{array}{cc}
	\zeta^{s} & \\
	& \zeta^{s} \\
	\end{array}%
	}%
	& & & & \\
	\cline{2-3}
	& \Rdash{} & {\quad} &  & & \\
	& & & \ddots  & &  \\
	& & & & {\quad} & \Ldash{} \\
	\cline{5-6}
	& & & & & \Ldash{%
	\begin{array}{cc}
	\zeta^{s\cdot(\Gamma-1)} & \\
	& \zeta^{s\cdot(\Gamma-1)} \\
	\end{array}%
	} \\
	\end{array}\right)\,.
\ee
From the discussion of gauge symmetry breaking in section \ref{sec:general-features}, the surviving gluons are simply
\be
\begin{split}
	&A_{A}^{\phantom{A}\,B} 
	= (A^{(\alpha)})_{a}^{\phantom{a}\, b}\,\delta_{\alpha}^{\phantom{\alpha}\beta} \qquad \alpha,\beta=0,\dots,\Gamma-1\quad a,b=1,2\,,\\
	&\widetilde A_{C}^{\phantom{A}\,D} 
	= (\widetilde A^{(\gamma)})_{c}^{\phantom{a}\, d}\,\delta_{\gamma}^{\phantom{\alpha}\delta} \qquad \gamma,\delta=0,\dots,\Gamma-1\quad c,d=1, 2, 3\,,
\end{split}
\ee
and the corresponding gauge symmetry of the daughter theory reads%
\footnote{Recall the presence of extra $U(1)$ factors in (\ref{eq:symmetry-breaking}). We take them to be lifted by means of the Stueckelberg mechanism \cite{Stueckelberg:1900zz}, see section \ref{sec:UV-completions} below.
}
\be
	\left(\prod_{\alpha=0}^{\Gamma-1} SU(2)^{(\alpha)} \right)\, \times\, \left( \prod_{\gamma=0}^{\Gamma-1} SU(3)^{(\gamma)} \right)\,.
\ee  
Matter fields are all bifundamentals, transforming under $\G$ as
\begin{align}
H_{A}^{\phantom{A}M} &\quad \mapsto\quad \big(\gamma_{F}^{s\,\,*}\big)^{M}_{\phantom{A}N}\,   \big(\gamma_{2}^{s}\big)_{A}^{\phantom{A}B}\,H_{B}^{\phantom{A}N} \,,\label{eq:ZgammaHtransfo}\\
Q_{C}^{\phantom{B}A}&\quad \mapsto\quad  (\gamma_{2}^{s\,\,*})^{A}_{\phantom{A}B}\,  (\gamma_{3}^{s})_{C}^{\phantom{A}D} \,Q_{D}^{\phantom{B}B}\,,\\
U_{M}^{\phantom{B}C}&\quad \mapsto\quad (\gamma_{3}^{s\,\,*})^{C}_{\phantom{A}D}\,  (\gamma_{F}^{s})_{M}^{\phantom{A}N} \,U_{N}^{\phantom{D}D}\,.
\end{align}
where $\gamma_{2}^{s},\, \gamma_{3}^{s}$ and $\gamma_{F}^{s}$ denote the regular embeddings of $\G$ into $SU(2\Gamma), \, SU(3\Gamma)$ and $SU(\Gamma)$ respectively, and lower indices in matter fields are  fundamental indices, while upper ones correspond to anti-fundamentals.
The invariant components follow directly from (\ref{eq:bifund-invt})
\be\label{eq:abelian-invt-matter}
\begin{split}
	(h^{(\alpha)})_{a} &= H_{(\alpha,a)}^{\phantom{\alpha,a)}(\alpha,m)}\,\qquad \alpha=0,\dots,\Gamma-1\,,\quad a=1,2\,,\quad m=1\,,\\
	(q^{(\alpha)})^{\phantom{a}a}_{c} &= Q^{\phantom{(\alpha,a)}(\alpha,a)}_{(\alpha,c)}\,\qquad \alpha=0,\dots,\Gamma-1\,,\quad a=1,2\,,\quad c=1,2,3\,, \\
	(u^{(\gamma)})^{c}& = U^{\phantom{(\alpha,a)}(\gamma,c)}_{(\gamma,m)}\,\qquad \gamma=0,\dots,\Gamma-1\,,\quad m=1\,,\quad c=1,2,3\,.
\end{split}
\ee

As a small aside, it is also instructive to take a look at explicit calculations, without relying on (\ref{eq:bifund-invt}); these turn out to be rather easy in the example at hand. For example, from (\ref{eq:Zgammarep}) and (\ref{eq:ZgammaHtransfo}) the explicit transformation of the Higgs multiplet $H_{A}^{\phantom{A}M}$ reads simply
\be\label{eq:explicit-Zgamma-action}
\begin{split}
	\zeta^{s}\,: H_{A}^{\phantom{A}M} %
	& \, \mapsto \, \big(\zeta^{s\cdot \alpha}\, \delta_{\alpha}^{\phantom{\alpha}\beta}\, \delta_{a}^{\phantom{\alpha}b}\big)\,\big(\zeta^{-s\cdot \mu}\, \delta^{\mu}_{\phantom{\alpha}\nu}\, \delta^{m}_{\phantom{\alpha}n}\big)\, H_{(\beta,b)}^{\phantom{(\beta,b)}(\nu,n)}
\end{split}
\ee
from which one sees immediately that invariant components are indeed those with $\alpha = \mu$ (mod $\Gamma$), consistently with the prediction (\ref{eq:abelian-invt-matter}).

\begin{figure}[h!]
\begin{center}
\includegraphics[width=1.0\textwidth]{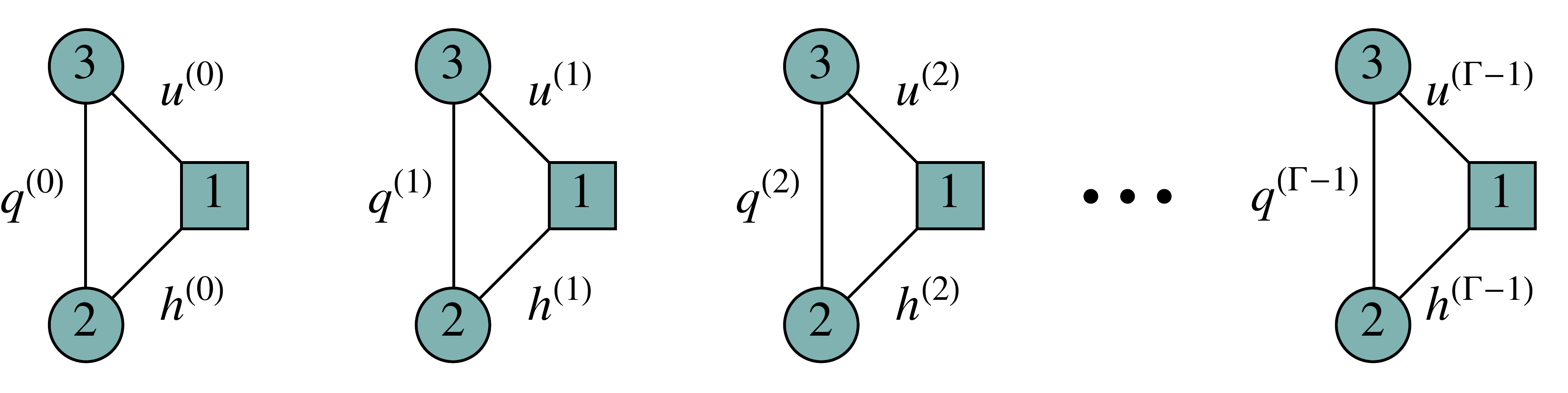}
\caption{The daughter quiver theory of the $\IZ_{\Gamma}$ orbifold. Nodes are blind to each other with respect to tree level gauge interactions, recall however that there will be cross-nodal quartic interactions for the Higgs fields, descending from the mother theory quartic term. The SU(1) flavor node is trivial, and is added to the figure for completeness only.
}
\label{fig:abelianquiver}
\end{center}
\end{figure}

What kind of interacting structure does the daughter theory have? A glance at (\ref{eq:abelian-invt-matter}) reveals that $h^{(\alpha)}$ transforms as a doublet of $SU(2)^{(\alpha)}$ and a singlet of all other daughter gauge groups. Similarly $(q^{(\alpha)})^{\phantom{a}a}_{c} $ is in the $\two\otimes\three$ of  $SU(2)^{(\alpha)}\times SU(3)^{(\alpha)}$ and a singlet under everyone else, while $(u^{(\alpha)})^{c}$ is a $\overline\three$ of $SU(3)^{(\alpha)}$.
Overall we find a daughter theory with $\Gamma$ sectors: each sector has gauge symmetry  $SU(2)^{(\alpha)}\times SU(3)^{(\alpha)}$, together with matter consisting of $h^{(\alpha)},\, q^{(\alpha)},\, u^{(\alpha)}$ in the corresponding (bi)-fundamental representations. 
The structure of this daughter theory is conveniently summarized by the quiver diagram of figure \ref{fig:abelianquiver}, and fits nicely with the discussion below equation (\ref{eq:bifund-invt}). 

It furthermore follows from (\ref{eq:abelian-invt-matter}) that Yukawa and gauge couplings in the mother theory generate corresponding interactions in the daughter theory, privately within each sector. 
In $\mathbb{Z}_\Gamma$ case the natural rescaling of couplings is trivial since all irreps have dimension one.

The quiver structure exhibits a discrete $S_{\Gamma}$ symmetry, acting on the quiver by permutations of the nodes. This clearly coincides with $\CS_{\vec d}$ from section \ref{sec:discrete}, and is therefore also a symmetry of the potential. 
We recognize $\CS_{\vec d} = S_{\Gamma}$ as the direct generalization of the $\mathbb{Z}_2$ at the heart of the Twin Higgs (clearly, $S_{2}\simeq \IZ_{2}$). 
As in the Twin Higgs, the role of $S_{\Gamma}$ can be understood from a bottom-up perspective: its main role is to ensure that gauge and yukawa couplings of all nodes are actually equal
\be
y^{(\alpha)}=y^{(\beta)}\,,\qquad  g^{(\alpha)}=g^{(\beta)}\,,\qquad \forall\, \alpha,\beta
\ee
This is both necessary and sufficient to ensure the radiative stability of the accidental $SU(2\Gamma)$ in the quadratic action:  
\begin{eqnarray} \label{eq:cws3}
V^{(1)} \supset \frac{\Lambda^2}{16 \pi^2} \left( - 6 y^2 + \frac{9}{4} g_2^2 + (4 \Gamma + 2) \lambda \right)  \left( \sum_{\alpha=0}^{\Gamma-1} |h^{(\alpha)}|^2 \right)\,.
\end{eqnarray}
While in the Twin Higgs the crucial discrete symmetry was something of an ad-hoc assumption,
it is just a natural feature of the Orbifold Higgs!

\subsection{Nonabelian Example: $S_3$}\label{sec:S3-toy}
We now turn to the more involved non-abelian orbifolds. Here we analyze the simplest case of $S_{3}$, but our framework extends straightforwardly to other discrete non-abelian groups. Another example, involving $\G = A_{4}$, has been relegated to appendix \ref{app:A4-toy}. 

Let us start by summarizing some basic facts about $S_{3}$ and its representations. The group elements read
\be
	g_{0}=e,\quad g_{1}=(1,2,3),\quad g_{2}=(3,2,1),\quad g_{3} = (1,2),\quad g_{4}=(2,3),\quad g_{5} = (3,1)\,.
\ee
We will denote the three irreps
\be
\begin{array}{ccl}
	r_{0} \,:& \quad d_{0} =1 \quad &\quad  \text{trivial irrep} \\
	r_{1} \,:& \quad d_{1} =1 \quad &\quad  \text{sign irrep} \\
	r_{2} \,:& \quad d_{2} =2 \quad &\quad  2\times2\; \text{matrices}\,,
\end{array}
\ee
in terms of which the {regular} representation decomposes as follows
\be
	\gamma^{s}=r^{s}_{0}\oplus r^{s}_{1}\oplus r^{s}_{2}\oplus r^{s}_{2}\,.
\ee
For additional clarity, the explicit form of the regular embedding of $\G=S_{3}$ into $SU(12)$ reads
\be
	\gamma_{N=2}^{s} = \left(\begin{array}{cccc}
	\begin{array}{cc}
	r_{0}^{s} & \\
	& r_{0}^{s}
	\end{array}%
	& \Ldash{} & &\\
	\cline{1-2}
	& %
	\LRdash{%
	\begin{array}{cc}
	r_{1}^{s} & \\
	& r_{1}^{s} \\
	\end{array}%
	}%
	& & \\
	\cline{2-4}
	& \Rdash{} &{%
	\Big(\ r_{2}^{s} \ \Big)_{2\times 2}%
	} & \\
	& \Rdash{} & & {%
	\Big(\ r_{2}^{s} \ \Big)_{2\times 2}%
	} 
	\end{array}\right)\,,
\ee
with a similar formula holding for $SU(18)$.

The mother theory is once again summarized by table \ref{tab:mothertheory}, where now 
$\Gamma=6$ is fixed.
Following (\ref{eq:symmetry-breaking}), the pattern of symmetry breaking is 
\be
\begin{split}
	SU(12) &\quad\longrightarrow \quad SU(2)^{(0)}\,\times \, SU(2)^{(1)}\,\times \, SU(4)^{(2)}\\
	SU(18)&\quad\longrightarrow \quad SU(3)^{(0)}\,\times \, SU(3)^{(1)}\,\times \, SU(6)^{(2)} \\
	SU(6)&\quad\longrightarrow \quad SU(2)^{(2)}\quad \text{(flavor)} \, .
\end{split}
\ee
Note in particular that we have a residual $SU(2)$ flavor symmetry, unlike in the abelian case.
For the sake of clarity, $SU(12)$ multi-indices in this context take the following values:
\be
	A=(\alpha,a,\a):\quad \alpha=0,1,2\quad %
	a=\left\{\begin{array}{l}  %
	1,2\ \ \text{for} \ \alpha=0,1 \\%
	1,\dots,4\ \ \text{for} \ \alpha=2 %
	\end{array}\right.\quad %
	\a=\left\{\begin{array}{l}  %
	0\ \ \text{for} \  \alpha=0,1 \\%
	0,1\ \ \text{for} \ \alpha=2 %
	\end{array}\right.%
\ee
\noindent The $SU(12)$ gauge fields $({A^{i}})_{A}^{\phantom{A}B}$ surviving the orbifold are the block-diagonal ones
\be
\begin{split}
	SU(2)^{(0)} &:\quad ({A^{(0),i_{0}}})_{(0,a,0)}^{\phantom{(0,a,0)}{(0,b,0)}}\qquad a,b=1,\dots,2 \quad i_{0}=1,\dots,3\\
	SU(2)^{(1)} &:\quad ({A^{(1),i_{1}}})_{(1,a,0)}^{\phantom{(0,a,0)}{(1,b,0)}}\qquad a,b=1,\dots,2 \quad i_{1}=1,\dots,3\\
	SU(4)^{(2)} &:\quad ({A^{(2),i_{2}}})_{(2,a,\a)}^{\phantom{(0,a,0)}{(2,b,\b)}}\qquad a,b=1,\dots,4\quad \a,\b=0,1 \quad i_{2}=1,\dots,15
\end{split}
\ee
Schur's lemma requires the gauge fields of $SU(4)^{(2)}$ to be of the form
\be
	({A^{(\alpha),i}})_{(2,a,\a)}^{\phantom{(0,a,0)}{(2,b,\b)}} = ({A^{(\alpha),i}})_{(2,a)}^{\phantom{(0,a}{(2,b)}}\,\delta_{\a}^{\phantom{\a}\b}\,.
\ee
Analogous statements apply to the surviving gluons of $SU(18)$. 

Turning to the flavor symmetry,  the $\G$-action on the $SU(6)$-flavor fundamentals reads simply
\be
\begin{split}
	& (\gamma_{F}^{s})_{M}^{\phantom{M}N} = (\mathbb{1}_{F}\otimes \gamma_{}^{s})_{M}^{\phantom{M}N} = \delta_{\mu}^{\phantom{\mu}\nu}\delta_{m}^{\phantom{m}n}\, (r^{s}_{\mu})_{\m}^{\phantom{\m}\n} 
	\\
	M = (\mu,m,\m): \quad & \mu=0,1,2\quad %
	m=\left\{\begin{array}{l}  %
	1\ \ \text{for} \ \mu=0,1 \\%
	1,2\ \ \text{for} \ \mu=2 %
	\end{array}\right.\quad %
	\m=\left\{\begin{array}{l}  %
	0\ \ \text{for} \  \mu=0,1 \\%
	0,1\ \ \text{for} \ \mu=2 %
	\end{array}\right.%
\end{split}
\ee

We should then analyze projections of the matter fields. Starting with $H$, it transforms under $g_{s}\in S_{3}$ as $H_{A}^{\phantom{A}M}\,\mapsto\, (\gamma^{s}_{2})_{A}^{\phantom{A} B}\ (\gamma^{s\,\, *}_{F})^{M}_{\phantom{A} N}\ H_{B}^{\phantom{A}N}$. Relying on (\ref{eq:bifund-invt}), all that remains to be done is to examine the invariant fields case by case, i.e. for $\alpha=0,1,2$.

\noindent{\underline{$\alpha=\mu=0$}}
\be
	d_{\alpha}=d_{\mu}=1\quad\Rightarrow\quad \a,\m=0, \quad a=1,2,\quad m=1,
\ee
so we are left with
\be
	(h^{(0)})_{a}\,=\,H_{(0,a,0)}^{\phantom{(0,a,0)}(0,m,0)}\qquad\text{in the}\ \Box\ \text{of}\ SU(2)^{(0)}\,,
\ee
we suppressed the flavor index on the LHS since it is fixed to $m=1$.

\noindent{\underline{$\alpha=\mu=1$}}
\be
\begin{split}
	d_{\alpha}=d_{\mu}=1\quad& \Rightarrow\quad \a,\m=0, \quad a=1,2,\quad m=1,
\end{split}
\ee
so we are left with
\be
	(h^{(1)})_{a}\,=\,H_{(1,a,0)}^{\phantom{(0,a,0)}(1,m,0)}\qquad\text{in the}\ \Box\ \text{of}\ SU(2)^{(1)}\,,
\ee

\begin{figure}[h!]
\begin{center}
\includegraphics[width=0.6\textwidth]{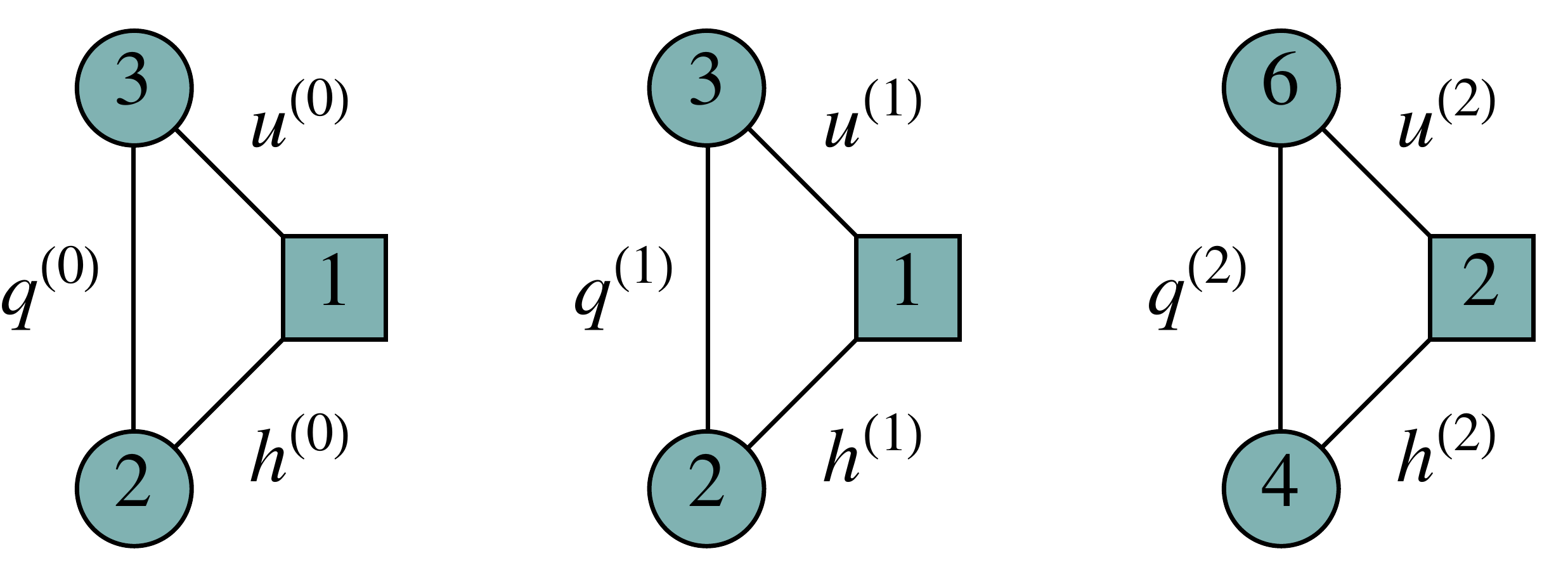} 
\caption{The daughter quiver theory of the $S_{3}$ orbifold. The flavor nodes for the first two sectors are trivial. Sectors are blind to each other with respect to tree level gauge interactions, recall however that there will be cross-nodal quartic interactions for the Higgs fields, descending from the mother theory quartic term.}
\label{fig:nonabelianquiver}
\end{center}
\end{figure}

\noindent{\underline{$\alpha=\mu=2$}}
\be
\begin{split}
	d_{\alpha}=d_{\mu}=2\quad  \Rightarrow\quad \a,\m=0,1, \quad & \quad a=1,\dots,4,\quad m=1,2,
\end{split}
\ee
therefore the invariant combination is
\be
	(h^{(2)})_{a}^{\phantom{a}m} \,=\,{1\over \sqrt{2}}\left( H_{(2,a,0)}^{\phantom{(2,m,0)}(2,m,0)}+ H_{(2,a,1)}^{\phantom{(2,m,0)}(2,m,1)}\right)\,,
\ee
in the $\Box\times\overline\Box$ of $SU(4)^{(2)}\, \times \, SU(2)^{(2)}$. As expected from our general discussion of the bifundamental orbifold (cf. figure \ref{fig:simplequiver}), we find a nontrivial flavor symmetry on the $\alpha=2$ node (in the case at hand $F\cdot d_{2} = 2$), indeed this exotic node the Higgs sector consists of \emph{a flavor doublet of quadruplets}!
The analysis for the $Q$ and $U$ fields is completely analogous and the full matter content of the daughter is summarized by table \ref{tab:s3daughter} and figure \ref{fig:nonabelianquiver}. 
Just like in the abelian case of section \ref{sec:abelian-toy}, the Yukawa and gauge couplings in the mother theory generate corresponding interactions in the daughter theory, privately within each sector.

\begin{table}[h!]
\begin{equation*}
 \resizebox{1\hsize}{!}{$\displaystyle 
 \begin{array}{c|c|cc|cc|cc|c}
	& \text{mother theory d.o.f.} & SU(2)^{(0)}& SU(3)^{(0)} & SU(2)^{(1)}& SU(3)^{(1)} &SU(4)^{(2)}& SU(6)^{(2)}&SU(2)^{(2)} \\
	\hline%
	h^{(0)} & H_{(0,a,0)}^{\phantom{(0,m,0)}(0,m,0)} &\Box & 1 & 1&1  & 1 & 1 & 1 \\
	h^{(1)} & H_{(1,a,0)}^{\phantom{(0,m,0)}(1,m,0)} & 1& 1&\Box &1  & 1 & 1 & 1 \\
       h^{(2)} & {1\over \sqrt{2}}\left( H_{(2,a,0)}^{\phantom{(0,m,0)}(2,m,0)}+ H_{(2,a,1)}^{\phantom{(0,m,0)}(2,m,1)}\right)&1 & 1 &1  & 1 & \Box& 1 & \overline\square \\\cdashline{1-9}
	q^{(0)} & Q^{\phantom{(0,a,0)}(0,a,0)}_{(0,c,0)} & \overline\square& \square & 1 & 1&1 & 1&1 \\
	q^{(1)} & Q^{\phantom{(0,a,0)}(1,a,0)}_{(1,c,0)}  & 1 & 1&\overline\square&\square & 1&1&1 \\
	q^{(2)} & {1\over \sqrt{2}}\left( Q^{\phantom{(0,a,0)}(2,a,0)}_{(2,c,0)}+ Q^{\phantom{(0,a,0)}(2,a,1)}_{(2,c,1)}\right)  & 1 & 1& 1 & 1 & \overline\square&\square &1 \\\cdashline{1-9}
	u^{(0)} & U^{\phantom{(0,a,0)}(0,c,0)}_{(0,m,0)} &1 & \overline\square & 1&1  & 1 & 1 & 1 \\
	u^{(1)} & U^{\phantom{(0,a,0)}(1,c,0)}_{(1,m,0)}  & 1&1&1 & \overline\square   & 1 & 1 & 1 \\
	u^{(2)} & {1\over \sqrt{2}}\left( U^{\phantom{(0,a,0)}(2,c,0)}_{(2,m,0)}+ U^{\phantom{(0,a,0)}(2,c,1)}_{(2,m,1)}\right)  & 1 & 1& 1&1&1 & \overline\square   & \Box \\
\end{array}$
}
\end{equation*}
\caption{Matter content of the daughter theory for the $S_3$ orbifold.\label{tab:s3daughter}}
\end{table}

Before turning to the symmetry properties of the daughter theory, let us provide a simple check of our results.
Rather than presenting the explicit computation of the matrix elements of projectors (a straightforward but unenlightening exercise), we will be content with a simpler, partial check. The Clebsch-Gordan decomposition for the $S_3$ irreps can be obtained by means of standard representation theory, and reads
\be\label{eq:CG-table-S3}
\begin{split}
	& r_{0}\otimes r_{\alpha} = r_{\alpha}\qquad \alpha=0,1,2 \\
	& r_{1}\otimes r_{1} = r_{0}\qquad r_{1}\otimes r_{2}=r_{2} \\
	& r_{2}\otimes r_{2} = r_{0}\oplus r_{1}\oplus r_{2}\,.
\end{split}
\ee
This provides a simple way of counting the number of surviving Higgses: the tensor product of gauge and flavor $S_{3}$-representations decomposes into
\be
\begin{split}
	\gamma_{2}\otimes \gamma_{F} & =\mathbb{1}_{2}\otimes \gamma\otimes\gamma \\
	& = \mathbb{1}_{2}\otimes \big( \mathbb{1}_{6}\otimes  r_{0}  \oplus \mathbb{1}_{6} \otimes  r_{1} \oplus \mathbb{1}_{12} \otimes r_{2} \big)
\end{split}
\ee
meaning that the orbifold should preserve $6 \cdot 2$ Higgs degrees of freedom, which correctly account for all $12$ components predicted from our application of (\ref{eq:bifund-invt}) above.

Just like in the $\IZ_\Gamma$ case, from the quiver structure in figure \ref{fig:nonabelianquiver} we can immediately read off both the exact discrete symmetry $\CS_{\vec d}$ and the accidental continuous symmetry $\CS_{acc}$: the former is now  the $\IZ_{2}$ from permuting the $\alpha=0,1$ sectors, the latter is an SU(12) rotating the $h^{(\alpha)}$. However in contrast to the $\mathbb{Z}_\Gamma$ orbifold, the discrete symmetry alone is not sufficient the guarantee the radiative stability of the full $SU(12)$ accidental symmetry of the quadratic action. Here the rescaling of the coupling constants, as dictated by the orbifold, plays a crucial role. In the case of the $S_3$ example, the couplings rescale as
\be
\begin{split}
y&=y^{(0)}=y^{(1)}\quad\text{and}\quad y^{(2)}=\frac{y}{\sqrt{2}}\\
g&=g^{(0)}=g^{(1)}\quad\text{and}\quad g^{(2)}=\frac{g}{\sqrt{2}} \quad \text{(for both gauge groups)}.
\end{split}
\ee
Recall from the discussion of section \ref{sec:accidentalsym} that the rescaling of the gauge couplings is just right to ensure that the SU(12) accidental symmetry of the one-loop quadratic action is only broken by a parametrically small amount. A similar argument holds for the yukawa couplings
\be
\begin{split}\label{eq:topeffaction}
V^{(1)}\supset& -\frac{ 3}{8 \pi^2} y^2  \Lambda^2 \sum_{\alpha=0}^1|h^{(\alpha)}|^2-\frac{ 3\times d_\alpha}{8 \pi^2}  \left(\frac{y}{\sqrt{d_\alpha}}\right)^2  \Lambda^2 |h^{(2)}|^2\\
&=-\frac{ 3}{8 \pi^2} y^2  \Lambda^2 \sum_{\alpha=0}^2|h^{(\alpha)}|^2.
\end{split}
\ee 
Observe that the one loop correction from the top yukawa \emph{is the same for all sectors of the quiver}, regardless of their size $d_{\alpha}$. 
The last line in (\ref{eq:topeffaction}) is therefore manifestly $SU(12)$ symmetric%
\footnote{A similar argument could be made for the single trace quartic, if one were to include it in the model.}. 
With (\ref{eq:topeffaction}) taken into account, the full quadratic one loop effective action reads 
\begin{eqnarray} \label{eq:cws3}
V^{(1)} \supset \frac{1}{16 \pi^2}\left( - 6 y^2 + \frac{9}{4} g_2^2 +26 \lambda \right)  \Lambda^2 \left( \sum_{\alpha=0}^2 |h^{(\alpha)}|^2 \right)
+ \frac{9 g_2^2 }{256 \pi^2}\Lambda^2|h^{(2)}|^2
\end{eqnarray}
manifestly exhibiting the $SU(12)$ symmetry, up to the last term which is parametrically small and acceptable for a cutoff in the tens of TeV. As such, in the $S_3$ orbifold Higgs model it is fair to identify the Higgs as a pseudo goldstone boson of the full approximate accidental $SU(12)$ symmetry.

\subsection{The fundamental role of bifundamentals}\label{sec:bifundamentals}
Before proceeding towards more complete models, we briefly comment on the essential role of the flavor symmetry. Consider the abelian case for instance: had we not included the flavor symmetry for the Higgs field, we would have got only $H_{A} = H_{(0,a)}$ surviving in the daughter theory. This would not only spoil the $S_{\Gamma}$ symmetry, but would also leave us with far fewer degrees of freedom than needed for generalizing the twin Higgs model. Conversely, had we included a third symmetry group acting nontrivially on $H$, we would have introduced extra unwanted degrees of freedom in the daughter theory. To illustrate this, notice that a straightforward generalization of  (\ref{eq:explicit-Zgamma-action}) would read
\be
\begin{split}
	g_{s}\,: H_{A}^{\phantom{A}MR} & \, \mapsto \, \zeta^{s\cdot (\alpha-\mu-\rho)}\,H_{A}^{\phantom{A}MR}\,,
\end{split}
\ee
such that the surviving fields would manifestly be those with $\alpha = \mu+\rho $ (mod $\Gamma$). We would then be left with $\Gamma$ Higgs doublets for each node of the daughter quiver, instead of just a single one. 

These considerations -- just direct consequences of standard representation theory -- clearly carry over to all matter fields of the mother theory. Hence, generally speaking, in order to retain a number of degrees of freedom that is suitable for model-building, the mother theory matter fields should all transform in \emph{bifundamentals} of those gauge or flavor groups on which $\G$ acts via the regular embedding.  This fact carries important consequences for models of phenomenological interest and we will return to this below in section \ref{sec:UV-completions}. 
We stress that, for the purpose of orbifolding matter fields, the distinction between flavor and gauge symmetries is immaterial. In fact, in the context of grand unified theories we will show that the flavor symmetry exploited in our toy models should be replaced by a gauge symmetry.

We hope to have convinced the reader, by means of the analysis in sections \ref{sec:abelian-toy} and \ref{sec:S3-toy}, of just how straightforward it is to ``carry out'' orbifolds in the regular representation with bifundamental matter. 
Nevertheless, a generalization beyond bifundamentals -- while probably uninteresting for Orbifold Higgs models -- is clearly possible, and in principle straightforward to carry out: the main extra difficulty is that (\ref{projectorOpp}) would not be valid in general, but this is easily fixed by the explicit knowledge of the irreps of $\G$, which would allow the construction of explicit projectors.

\subsection{A model-builder's guide to the Orbifold Higgs}\label{sec:recipe}
Although straightforward applications of standard orbifold technology, some of the technicalities of the notation in section \ref{sec:general-features} may obscure the simple nature of Orbifold Higgs models.  Nevertheless the overall lessons are both general and simple, and in this section we recollect them into a practical model-builder's guide to Orbifold Higgs models.

Generally speaking, for any given $\G$ all that is needed to write down the daughter model is really the tuple $\{d_{\alpha}\}_{\alpha}$ of dimensionalities of $\G$-irreps. 
These can be usually obtained by means of standard representation theory, or simply from the math literature. Much of the structure of Orbifold Higgs models is encoded into this simple data, as follows
\begin{itemize}
\item the model has one sector for each $\G$-irrep
\item the $\alpha$-th sector includes $SU(2 d_\alpha)\times SU(3 d_\alpha)$ gauge nodes, as well as an $SU(d_{\alpha})$ flavor node
\item each sector contains a single bi-fundamental fermion ($q^{(\alpha)}$), together with a flavor $d_\alpha$-plet of scalars ($h^{(\alpha)}$) in the fundamental of $SU(2 d_\alpha)$, and  another flavor $d_\alpha$-plet of fermions ($u^{(\alpha)}$) in the anti-fundamental of $SU(3 d_\alpha)$
\item the gauge and yukawa couplings of each sector satisfy
\begin{equation}
\left(\frac{g^{(\alpha)}}{g^{(\beta)}}\right)^{2}=\frac{d_\beta}{d_\alpha}\quad\mathrm{and}\quad\left(\frac{y^{(\alpha)}}{y^{(\beta)}}\right)^{2}=\frac{d_\beta}{d_\alpha}\quad \forall \;\alpha,\beta
\end{equation}
\item the tree-level Higgs potential is (approximately) $SU(2\Gamma)$ symmetric, this follows from the orbifold picture -- with the caveat that the single trace quartic operator in (\ref{eq:mother-operators}) be subdominant or absent in the mother theory. 
\end{itemize} 
These basic properties then ensure that  the accidental symmetry is automatically preserved for the two-point function of the $h^{(\alpha)}$ at one loop. Moreover it is guaranteed that the model descends from an orbifold projection of a theory of the form shown in table \ref{tab:mothertheory}, providing a natural avenue to a possible UV completion. One last nice feature worth keeping in mind is that every discrete group $\mathcal{G}$ has at least one irrep of dimension one, namely the trivial representation: this means that every Orbifold Higgs model is guaranteed to contain at least one sector which can be identified with the Standard Model.  When the $h^{(\alpha)}$ acquire vacuum expectation values, the Standard Model-like Higgs may then be identified as a pseudo-goldstone boson of the spontaneously broken accidental symmetry.

Finally we would like to stress that, while the examples considered in sections \ref{sec:abelian-toy} and \ref{sec:S3-toy} are just the two simplest applications, the scope of applicability of our formalism clearly extends well beyond them, and we hope these techniques will prove useful in future investigations of related ideas. 

\section{A Standard Model-like Higgs} \label{sec:lsm}

Thus far we have demonstrated that orbifolds of a mother theory can give rise to daughter theories exhibiting an accidental symmetry of the collective Higgs sector. A parametrically light Standard Model-like Higgs arises as a pseudo-goldstone boson of the spontaneous breaking of this accidental symmetry, with the gauge bosons and fermions of the various daughter sectors playing the role of partner states protecting the weak scale. Although the general mechanism is clear, it is useful to see how orbifold Higgs theories can give rise to a parametrically light Standard Model-like Higgs in an explicit toy example. The physics of the $\mathcal{G} = \mathbb{Z}_2$ case is identical to the twin Higgs \cite{Chacko:2005pe}, so to manifest the general features of the orbifold Higgs, let us consider the case of  $\mathcal{G} = \mathbb{Z}_3$. 

\subsection{The $\mathbb{Z}_3$ vacuum}
As we saw in section 3, the parent theory consists of an $SU(9)\times SU(6)$ gauge theory with matter fields $H, Q, U$ transforming under the gauge group and an $SU(3)$ flavor symmetry as detailed in table \ref{tab:mothertheory}. The tree-level potential for this toy model is simply 
\begin{equation}
V_p =  - m^2 \,|H|^2 +\lambda (|H|^2)^2 + \delta\, {\rm Tr}[H^\dagger H H^\dagger H]  + y \,H Q U
\end{equation}
where we include both a single-trace quartic $\delta$ and a double-trace quartic $\lambda$, with the understanding that $\delta \ll \lambda$ for a viable model.

The daughter theory consists of an $ \prod_{\alpha=0}^2 SU(3)^{(\alpha)} \times SU(2)^{(\alpha)}$ gauge theory (again omitting $U(1)$ factors). Each gauge sector labeled by $\alpha$ contains a Higgs doublet $h^{(\alpha)}$, $SU(3)$-triplet/$SU(2)$-doublet quark $q^{(\alpha)}$, and $SU(3)$-triplet/$SU(2)$-singlet quark $u^{(\alpha)}$. The Standard Model weak and color sectors can be identified with, say, the $\alpha = 0$ copy. The Higgs potential is 
\begin{equation} \label{eq:z3dpot}
V_d = - m^2 \left(\sum_\alpha |h^{(\alpha)}|^2 \right) + \lambda \left( \sum_\alpha |h^{(\alpha)}|^2 \right)^2 + \delta \left(\sum_\alpha |h^{(\alpha)}|^4\right)  + y \left( \sum_\alpha h^{(\alpha)} q^{(\alpha)} u^{(\alpha)} \right)
\end{equation}
where we have taken care to group the fields in such a way as to emphasize the $SU(6)$-symmetric form of the mass terms and double-trace quartic $\lambda$. Of course, this is merely the tree-level quartic. As emphasized in section  \ref{sec:accidentalsym}, radiative corrections to the mass terms preserve the $SU(6)$-symmetric form,\footnote{We emphasize that radiative corrections to the mass proportional to $\lambda$ and $\delta$ {\it both} preserve the $SU(6)$ symmetry of the mass terms; as we will elaborate shortly, the smallness of $\delta$ is desired to keep the SM-like Higgs quartic sufficiently small, rather than due to radiative considerations.} while radiative corrections to the quartics contribute to both $\lambda$ and $\delta$. If $\delta \ll \lambda$ at tree level in the parent theory, these additional radiative corrections to $\delta$ remain adequately small provided a modest hierarchy between the cutoff and the weak scale.

The potential (\ref{eq:z3dpot}) possesses an absolute minimum with nonzero vevs for all three $h^{(\alpha)}$, namely $|\langle h^{(\alpha)} \rangle |^2 = m^2 / (3 \lambda + \delta)$. This higgses the accidental $SU(6)$ with order parameter $f^2 = \sum_\alpha |\langle h^{(\alpha)} \rangle|^2 = \frac{m^2}{\lambda + \delta / 3}$. The vevs also higgs all three $SU(2)$ gauge groups, such that nine of the twelve real degrees of freedom in the $h^{(\alpha)}$ are eaten. Of the remaining three states, one corresponds to the radial mode of spontaneous $SU(6)$ breaking, while the other two are uneaten pseudo-goldstones of the same spontaneous breaking. However, at this stage neither of the two goldstones may be identified with an SM-like Higgs; they are maximally-mixed linear combinations of components of $h^{(\alpha)}$ and, moreover, the weak scale is not well-separated from the scale of $SU(6)$ breaking since $\langle h^{(0)} \rangle = v = f / \sqrt{3}$. These properties are clearly expected from the full discrete $S_3$ symmetry of the daughter theory.

To obtain a realistic vacuum, the potential should be perturbed by terms that break the daughter $S_3$ (and hence also the accidental $SU(6)$). Such terms may be either soft (i.e.~mass terms) or hard (i.e.~quartic couplings). While these terms are incompatible with the parent symmetries, they may be introduced in geometric orbifolds as we will further discuss in the next section. Soft breaking terms have the virtue of preserving radiative stability at the cost of some amount of tree-level tuning, while hard breaking terms introduce additional radiative sensitivity to the cutoff but no associated tree-level tuning.

For simplicity, let us consider breaking the $S_3$ and $SU(6)$ by soft terms, which we choose to be of the form $\Delta V = \rho^2 \left(|h^{(0)}|^2 - \frac{1}{2} |h^{(1)}|^2 - \frac{1}{2} |h^{(2)}|^2 \right)$. The specific form is chosen merely to simplify expressions; any general $S_3$-breaking mass terms would suffice. Under this perturbation the vevs are
\begin{equation}
v^2 \equiv v_0^2 = \frac{m^2}{3 \lambda + \delta} - \frac{\rho^2}{\delta} \hspace{1cm} v_1^2 = v_2^2 = \frac{m^2}{3 \lambda + \delta} + \frac{\rho^2}{2 \delta}
\end{equation}
while $f$ has been kept fixed. Now arbitrary parametric separation between the weak scale and $SU(6)$ breaking, $v \ll f$, can be achieved by tuning $\rho^2/\delta$ against $m^2 / 3 \lambda$, a tree-level tuning of order $3 v^2 / f^2$. As we will see, this is ultimately not a prohibitive tuning for realistic theories, and indeed grows increasingly modest with larger values of $\Gamma$. Expanding in terms of $v/f$, there is now a goldstone mode $h \sim - \left(1 - \frac{v^2}{2f^2} \right) h^{(0)}_0 + \frac{1}{\sqrt{2}} \frac{v}{f} h^{(1)}_0 + \frac{1}{\sqrt{2}} \frac{v}{f} h^{(2)}_0$ of mass $m_h^2 \sim 3 \delta v^2$. This state may be identified with the Standard Model-like Higgs, as it is aligned with the weak vev up to $\mathcal{O}(v^2/f^2)$ corrections. Its mass and quartic are of suitable size provided $\delta \sim \lambda_{SM}$, i.e., the radiatively-corrected single-trace quartic is of the order of the Standard Model Higgs quartic.
 This is the reason for requiring $\delta$ to be small; if $\delta$ were of order the double-trace quartic $\lambda$, the $SU(6)$ would be badly broken and this would be reflected in a large quartic coupling for $h$. The remaining goldstone is $h' \sim - \frac{1}{\sqrt{2}} h_0^{(1)} + \frac{1}{\sqrt{2}} h_0^{(2)}$ with mass $m_{h'}^2 \sim  \delta f^2$, while the radial mode is $h_R \sim \frac{v}{f} h_0^{(0)} + \frac{1}{\sqrt{2}} h_0^{(1)} + \frac{1}{\sqrt{2}} h_0^{(2)}$ with mass $m_{h_R}^2 \sim 2 \lambda f^2$. Thus we arrive at a theory with a parametrically light SM-like pseudo-goldstone Higgs, a second pseudo-goldstone mode primarily aligned with the non-SM weak sectors and heavier by an amount $\propto f/v$, and a heavy radial mode.

As in any global symmetry scheme for a light Higgs, the maximum value of $v$ relative to $f$ is set by phenomenological considerations. The SM-like Higgs is rotated away from alignment with the SM vacuum expectation value by an amount $\sim v^2/f^2$, leading to Higgs coupling deviations of the same order. The collective $\mathcal{O}(10\%)$ precision of LHC Higgs coupling measurement suggests $v^2 / f^2 \sim 0.1$. If the smallness of $v$ is explained by soft breaking of the $S_3$ as described above, this corresponds to a modest $\sim 30\%$ tuning in the potential -- exceedingly natural by current standards.

The sense in which the gauge bosons and fermions of the non-SM sectors play the role of conventional partner particles despite not carrying Standard Model quantum numbers can be made explicit by integrating out the radial mode and studying the induced couplings of these states to the SM-like Higgs. Integrating out $h_R$ gives rise, for example, to interactions of the form
\begin{equation}
\mathcal{L} \supset - \frac{1}{2} y f q_0^{(1)} u^{(1)} - \frac{1}{2} y f q_0^{(2)} u^{(2)} + \frac{y^2 h^2}{4 y f}  q_0^{(1)} u^{(1)} +  \frac{y^2 h^2}{4 y f}  q_0^{(2)} u^{(2)}
\end{equation}
precisely the form of top partner couplings expected from a pseudo-goldstone Higgs, with the wrinkle that the role of top partner is shared in part by top quarks of two different sectors. 

\subsection{Generalizations: $\mathbb{Z}_\Gamma$ and $S_3$}
Although we have focused here on $\mathcal{G} = \mathbb{Z}_3$, the generalization to $ \mathbb{Z}_\Gamma$ is straightforward. There are $\Gamma - 1$ uneaten pseudo-goldstone states; soft or hard breaking of the $S_\Gamma$ allows $v \ll f$ and one pseudo-goldstone to be identified with the SM-like Higgs. In the case of soft breaking, the tree-level tuning required for $v \ll f$ decreases with increasing $\Gamma$ \cite{Foot:2006ru}, though even for $\Gamma = 3$ the apparent tuning is nominal. The remaining $\Gamma - 2$ pseudo-goldstones are heavier by an amount $f/v$ and mostly aligned with non-SM weak sectors, while the roles of partner states are played by admixtures of states in the various non-SM sectors.

More exotic cases such as $\mathcal{G} = S_3$ proceed along similar lines. Here there are modest radiative perturbations to the structure of the potential since there are three sectors but the exact discrete symmetry of the daughter theory is only $S_2$. This is reflected in the fact that radiative quartics in the $SU(4)$ sector are half as large as those in the $SU(2)$ sector. The relative rescaling of gauge and yukawa couplings for sectors of different size $d_\alpha$ ensures that radiative corrections to the quadratic potential (proportional to $y^2$ and $g^2$ at one loop) remain the same between sectors, but consequently implies that radiative corrections to the quartic potential (proportional to $y^4$ and $g^4$ at one loop) differ. Similarly, the $1/N$-suppressed mass corrections introduce additional perturbations to the potential. The quartic perturbation serves to align the $SU(12)$-breaking vev primarily with the $SU(4)$ gauge sector, while the mass perturbation may work in either direction depending on the sign of the cutoff $\Lambda^2$ for gauge loops and depends on the details of the UV completion. In any event, numerically the quartic dominates to give the $SU(4)$ gauge sector a larger share of the $SU(12)$-breaking order parameter than the $SU(2)$ gauge sectors. In contrast to the $ \mathbb{Z}_\Gamma$ case, the breaking of the approximate global $SU(12)$ symmetry does not lead to the complete breaking of the gauged $SU(4)$, but rather leaves behind an unbroken $SU(3)$ gauge group. This residual group may confine depending on the sign of the beta function or be higgsed upon confinement and chiral symmetry breaking in the accompanying $SU(6)$ color group. There are now ten uneaten pseudo-goldstones, of which eight combine to form an approximate two-Higgs-quadruplet model of the $SU(4)$ gauge sector. Without further perturbation of the potential, the remaining pair of pseudo-goldstones arrange into maximally-mixed combinations of states in the two $SU(2)$ gauge sectors. A further perturbation of the potential by hard or soft $S_2$-breaking terms results in one pseudo-goldstone mode being identified with the SM-like Higgs. As in the case of $ \mathbb{Z}_\Gamma$, the role of the top partner is played by a collection of top quarks in the non-SM sectors, albeit now including states charged under an $SU(6)$ color group! For these states the multiplicity and top yukawa scale precisely in tandem to cancel the quadratic sensitivity of the Standard Model top loop: the color trace now gives a factor of six, but the top yukawa-squared is scaled down by a factor of 2. {\it While this example may seem baroque, it provides -- to our knowledge -- the first concrete example of a natural theory based on symmetries where the relative multiplicity and coupling of top partners is scaled in such a way as to preserve cancellation of quadratic divergences at one loop.}

\section{UV Completions}\label{sec:UV-completions}

Up to this point we have focused on toy models to illustrate how the orbifold Higgs can protect the weak scale from radiative corrections related to the top and weak gauge sectors of the Standard Model. In this section we illustrate how these toy examples can be extended to complete models with all the features of the Standard Model. In general, any realistic model building effort necessarily breaks up into two parts:
\begin{enumerate}
\item Write down the `core' of the model which consists of the non-abelian gauge fields and $H$, $Q$ and $U$. As was discussed in section \ref{sec:examples}, this step is close to trivial and fully specified by the choice of orbifold group.
\item Dress the core model with the hypercharge interactions and the remaining Standard Model fields. Viable models typically include ingredients that do not transform as irreducible representations of the mother symmetry. From the four-dimensional perspective these correspond to introducing incomplete multiplets, which naturally arise in geometric orbifolds or deconstructions thereof. In this step the model builder will have to make a number of choices, as we will see in the remainder of this section.
\end{enumerate}
While detailed treatment of UV completions is beyond the scope of this work, here we will  discuss qualitative features of possible geometric constructions and summarize a few possible options for incorporating Standard Model hypercharge.

\subsection{Geometric orbifolds}\label{sec:geometric}
As explained in the introduction, the orbifold Higgs boils down to a linear $\sigma$-model where the Higgs is identified with a pseudo-goldstone boson of an accidental symmetry of the quadratic action. 
This linear $\sigma$-model should be UV completed at a scale $\Lambda$, where the full symmetry of the mother theory is assumed to be largely restored. Any hierarchy problems in the mother theory above the cutoff may be solved in the usual ways, such as supersymmetry or compositeness, without observable consequences below the cutoff. A straightforward way to obtain such a UV completion is through the geometric interpretation of the orbifold in a model with extra space-time dimensions \cite{Antoniadis:1990ew}. \footnote{For reviews on the subject, see for instance \cite{Ponton:2012bi,Csaki:2005vy}.} Not only does the geometric orbifold naturally provide a justification for the presence of the accidental symmetries of the low energy effective action, it also provides a mechanism for introducing incomplete multiplets of the mother symmetry through states localized on lower-dimensional defects of the higher-dimensional theory. This permits, for example, the introduction of first- and second-generation Standard Model fermions without corresponding partners in other sectors. Among other things, this alleviates the tension of the conventional twin Higgs with cosmology by reducing the number of light species.

Ultraviolet completions of $\mathbb{Z}_2$ orbifolds in terms of the dimensional reduction of a five-dimensional theory on an interval with non-trivial boundary conditions are well understood in the context of Scherk-Schwarz supersymmetry breaking \cite{Scherk:1978ta}
and GUT model building \cite{Kawamura:2000ir,Kawamura:2000ev,Hall:2001pg,Altarelli:2001qj,Hebecker:2001wq,Hebecker:2001jb}. {\it A priori} there are several ways in which orbifold Higgs models can be embedded in these UV completions. Perhaps the simplest possibility is to consider a single non-trivial $\mathbb{Z}_2$ which is enough to obtain the desired pattern of symmetry breaking in a non-supersymmetric context. In supersymmetric models such a setup is moreover sufficient break N=2 supersymmetry down to N=1, but some additional soft supersymmetry breaking is needed to break the remaining N=1 supersymmetry. In this case the low energy theories are generalizations of the supersymmetric twin Higgs \cite{Chang:2006ra, Craig:2013fga}.  Alternatively, one could consider traditional Scherk-Schwartz supersymmetry breaking, with non-trivial orbifolds on both boundaries, such that the N=2 supersymmetry is fully broken without the need for additional soft terms. In this case the gauge \& global symmetries may be broken on one or both boundaries, and the low energy theories are generalizations of folded supersymmetry \cite{Burdman:2006tz}. For simplicity, we set aside supersymmetry here and restrict ourselves to simplest case with a single non-trivial $\mathbb{Z}_2$ orbifold.  As was mentioned before, any fields which only fit in representations of the daughter but not in representations of the mother can be localized on the boundary where the symmetry is reduced. As an alternative to placing all fields with twin partners in the bulk, one could also consider placing some on the symmetric boundary. For instance, for the $H$ and $U$ fields such a setup no longer requires the additional global symmetry SU(2) symmetry, since all degrees of freedom of these boundary fields already appear in the low energy theory. An interesting consequence is that now the dreaded single trace quartic for $H$ cannot arise.

For genuine $\mathbb{Z}_\Gamma$ orbifolds (with $\Gamma>2$), however, there is no geometric action on $\mathbb{R}^{1,3}\times S^1$ yielding fixed points, and one is forced to consider at least a six-dimensional setup, where  $\mathbb{Z}_\Gamma$ orbifolds can naturally arise from conical singularities.\footnote{See \cite{Hebecker:2001jb} for a discussion in the context of GUT's.} A simple possibility is to consider $\mathbb{R}^{1,3}\times S^2$ with $\IZ_{\Gamma}$ acting by rotations on $S^{2}$; the poles of the 2-sphere are then the fixed orbifold points. More generally, it may be possibe to realize $S_n$ orbifolds geometrically by exploiting the $S_{n}$ action on an $(n-1)$-sphere.\footnote{We thank Duccio Pappadopulo for bringing this to our attention.} Concretely, considering the unit sphere in $\IR^{n}$,
\begin{equation}
\sum_{i=1}^n x^2_i =1
\end{equation}
we may take the $S_n$ to act by a permutation on the ambient coordinates $x_i$. Again we find two antipodal orbifold points:
\begin{equation}
x_i = \frac{1}{\sqrt{n}}\quad\mathrm{and}\quad x_i = - \frac{1}{\sqrt{n}} \,, \quad \forall i\,.
\end{equation}

It is worth emphasizing that geometric completions provide a natural context for lifting any unwanted $U(1)$ gauge groups via the Stueckelberg mechanism, as alluded to in previous sections. This would consist of a generalized Green-Schwarz mechanism in a higher-dimensional theory, and may give mass to both anomalous and non-anomalous $U(1)$ gauge fields \cite{Ibanez:1998qp, Poppitz:1998dj, Ibanez:2001nd} (for a nice discussion, see \cite{Ghilencea:2002da}). 
When $U(1)$ factors are lifted in this way, they leave behind $U(1)$ global symmetries in the infrared, which in turn induce selection rules governing yukawa couplings. Note that in all the cases we have considered, the top yukawa is automatically compatible with such selection rules.
 
It is also possible to realize the physics of geometric orbifolds in a purely four-dimensional context through dimensional deconstruction \cite{ArkaniHamed:2001ca}. The bulk physics is reproduced by some number of sites possessing the gauged parent symmetry group, while defect physics is reproduced by a single site with only the daughter symmetries gauged. For example, the simple $[SU(6) \times SU(4)] /\mathbb{Z}_2$ orbifold can be reproduced by a simple two-site model where one site consists of an $SU(6) \times SU(4)$ gauge group and the other of an $[SU(3)\times SU(2)]^2$ gauge group, higgsed down to a diagonal $[SU(3)\times SU(2)]^2$ group by the vacuum expectation values of scalars transforming as bi-fundamentals of the two sites. Generically third generation fermions should be charged under the parent site, while first- and second-generation fermions can be charged under the daughter site. Provided the gauge couplings of the parent site are much smaller than the gauge couplings of the daughter site, the gauge couplings of the diagonal massless $[SU(3)\times SU(2)]^2$ gauge bosons will be largely inherited from the parent couplings. It is relatively straightforward to generalize this construction to other abelian orbifolds, including orbifold reduction of supersymmetry \cite{Craig:2014fka}. However, to our knowledge only abelian orbifolds have been reproduced by dimensional deconstruction; it should also be possible -- and quite interesting -- to deconstruct non-abelian orbifolds as well.

\subsection{Matter fields, Hypercharge and Unification}
Once we committed to a choice of orbifold group and geometry, what remains to be done is to add the hypercharge interactions and the remaining matter fields subject to considerations of anomaly cancellation.  For simplicity, in this section we restrict ourselves to the $\mathbb{Z}_2$ orbifold with the understanding that the key features can be readily generalized to generic $\G$
\footnote{With the obvious caveat that anomaly issues are quite different in $5$ and $6$ dimensions.}, %
in a rather straightforward manner, by means of the constructions discussed in previous sections. 

As mentioned above,
$\mathbb{Z}_2$ orbifolds enjoy a natural geometric realization on $\mathbb{R}^4\times S^1/\mathbb{Z}_2$ and such constructions are automatically anomaly free provided the 4D low energy theory is anomaly free \cite{ArkaniHamed:2001is}. To incorporate hypercharge interactions, first note that a $U(1)$ gauge symmetry of the mother theory would survive the orbifold intact (as is evident from (\ref{eq:gluon-orbifold}))
and therefore it follows that all fields in the daughter theory would be charged under it. This scenario is highly constrained by direct searches for states coupling through the $Z$; heavy $Z'$s; and heavy stable charged particles, and for simplicity we do not pursue this option any further. 

An attractive and viable alternative is for the hypercharge group to arise in the daughter theory from orbifolding a non-abelian gauge group (recall the extra $U(1)$ factors in (\ref{eq:symmetry-breaking}), which we have thus far neglected), 
in which case the dark sectors do not necessarily need to carry hypercharge quantum numbers. However in this scenario, the fact that Q must be a bifundamental (for the reasons discussed in section \ref{sec:bifundamentals}) demands that at least one of the generators contributing to hypercharge must be embedded in the non-abelian groups from which the color and weak isospin groups ultimately descend.\footnote{{ This non-abelian hypercharge embedding also has implications for possible kinetic mixing between $U(1)$ factors in the daughter theory, which is tightly constrained if additional $U(1)$ bosons remain massless. In general, low-scale embedding of $U(1)$ factors into non-abelian groups leads to vanishing kinetic mixing, though in UV complete models splittings in unified multiplets may give rise to suppressed contributions \cite{Dienes:1996zr}. While kinetic mixing is extremely small at the level of the daughter theories and may be rendered entirely innocuous by lifting additional $U(1)$ factors with the Stueckelberg mechanism, it would be interesting to study kinetic mixing in UV complete theories with low-scale unified embeddings to see if constraints are naturally satisfied for additional massless $U(1)$'s.}}
 In other words, the phenomenology of these models enforces (partial) gauge coupling unification! Interestingly, ultra low scale ($\sim 10$ TeV) gauge coupling unification may be achieved rather easily in higher dimensional models \cite{Dienes:1998vg,Dienes:1998vh}
In what follows, we merely assume gauge coupling unification at a scale below the cut-off of the 5D model. With this assumption, we sketch several possible ways by which a more complete model may be achieved, without attempting to provide an exhaustive list. In particular, we restrict ourselves to the `minimal' setup, where only the third generation is required to reside in the bulk, while the first two generations are confined on the Standard Model brane.

\subsubsection{$U(6)\times U(4)$}
 Consider a mother $U(6)\times U(4)$ gauge theory in the bulk, for which there are 4 $U(1)$'s surviving the orbifolding procedure: two come from decomposing the unitary groups into a product of the $U(1)$ with a special unitary group, while the other two $U(1)$'s are of the type in (\ref{eq:symmetry-breaking}). Concretely, in a suitable choice of basis the generators read
\begin{align}
T^{(6)}=\left(\!\!\begin{array}{cc} \mathbb{1}_{3}&\\&\mathbb{1}_{3} \end{array}\!\!\right)\quad\quad T'^{(6)}=\left(\!\!\begin{array}{cc} \mathbb{1}_{3}&\\&-\mathbb{1}_{3} \end{array}\!\!\right)\nonumber\\
T^{(4)}=\left(\!\!\begin{array}{cc} \mathbb{1}_{2}&\\&\mathbb{1}_{2} \end{array}\!\!\right)\quad\quad T'^{(4)}=\left(\!\!\begin{array}{cc} \mathbb{1}_{2}&\\&-\mathbb{1}_{2} \end{array}\!\!\right)
\end{align} 
where the subscripts on the generators denote which unitary group in the mother theory they descend from. Needless to say, they only act non-trivially on the subspace on which their corresponding gauge group in the mother theory acts. The following linear combination of generators reproduces the correct Standard Model hypercharge with zero charge for the fields in the twin sector:
\begin{equation}\label{uxuhypercharge}
Y=\frac{1}{3}(T^{(6)}+T'^{(6)})+\frac{1}{4}(T^{(4)}+T'^{(4)}).
\end{equation}
Note that to preserve charge quantization, we implicitly assumed gauge coupling unification in (\ref{uxuhypercharge}). The remaining linear combinations may be lifted with the Stueckelberg mechanism. Interestingly, in this model, the hypercharge assignments of all the bulk fields are fully fixed by their representations under the $U(6)\times U(4)$ in the bulk. It is therefore not possible to add the rest of the Standard Model to the bulk and we must localize these fields on the Standard Model brane. The Standard Model sector is anomaly free by construction, but the twin sector is not. This can be addressed by defect-localizing some spectator states charged under the dark gauge groups, without spoiling the one loop protection of the Higgs.

Finally, observe that the $T^{(6)}$ generator trivially commutes with all other generators, and plays the role of the baryon number in the mother theory. While Standard Model baryon number in the daughter theory is violated, $B+B'$ is preserved, with $B$ and $B'$ the baryon number in Standard Model and the twin sector respectively. This implies that neutron/anti-neutron oscillations are always forbidden. The proton on the other hand may decay, but only to a final state carrying hidden sector baryon number. Generically, we expect the lightest twin baryon to be heavier than the proton, in which case proton decay is kinematically forbidden. Even if this is not the case, the proton decay to the twin sector would occur through a dimension 12 operator, and since the proton degrees of freedom are confined on the brane, the decay would have to occur through mixing with the third generation. The high dimensionality of the operator in combination with the CKM suppression ensures that the bounds on invisible proton decay from the SNO experiment \cite{Ahmed:2003sy} are easily evaded, even if the dark baryons are lighter than the proton.

\subsubsection{Pati-Salam unification}
Another simple alternative is to consider a generalization of Pati-Salam unification \cite{Pati:1974yy}. Concretely, consider the orbifold
\begin{equation}
SU(8)\times SU(4)\times SU(4)/\mathbb{Z}_2 \rightarrow [SU(4)\times SU(2)\times SU(2) ]^2 \times U(1)^3.
\end{equation}
Each Pati-Salam factor in the daughter will then provide its own hypercharge gauge group after the usual breaking to $SU(3)\times SU(2)\times U(1)$, ensuring that the twin states are dark under the Standard Model hypercharge. Also here the additional $U(1)$ factors are lifted with a Stueckelberg mechanism. The matter content of a single generation plus twin particles fits in a
\be
	({\bf 8},\bar{\four},\one) \oplus (\bar{\bf 8},\one,\four) \rightarrow  ({\bf 4},\bar{\two},\one)_{\mathrm{sm}} \oplus (\bar{\bf 4},\one,\two)_{\mathrm{sm}}\oplus ({\bf 4},\bar{\two},\one)_{\mathrm{twin}} \oplus (\bar{\bf 4},\one,\two)_{\mathrm{twin}}
\ee
while two Higgs fields with opposite hypercharge plus their twin partners fit in
\be
(\one,\four,\bar{\four}) \rightarrow  (\one,\two,\bar{\two})_{\mathrm{sm}} \oplus  (\one,\two,\bar{\two})_{\mathrm{twin}}
\ee
In the Pati-Salam setup a two Higgs doublet model is therefore a natural possibility and in this case and all third generation Standard Model representations may be put in the bulk.  Both the twin and the SM sectors are then anomaly free by construction.

Although the Pati-Salam gauge sector does not commute with baryon number, the latter is nevertheless preserved in all low energy processes if the Pati-Salam symmetry is broken down to the Standard Model by a GUT orbifold mechanism. 
For the hidden sector decays, the argument from the $U(4)\times U(6)$ model can then be generalized to the effective baryon number of the Pati-Salam setup.\footnote{If the dark baryons are lighter than the proton and if one allows for more than one generation in the bulk, the proton decay rate may be enhanced to the extend that it could be experimentally accessible.}  For low scale Pati-Salam models, there furthermore are a number of interesting constraints from rare meson decays, which probe unification scales as high as $\sim 10^3$ TeV,  if all generations reside in the bulk. (For a recent update on these constraints, see \cite{Kuznetsov:2012ad} and references therein.) In the orbifold Higgs scenario it however suffices that the only the third generation resides in the bulk, in which case we expect the constraints on the unification scale to be substantially weaker. While we do not attempt it here, a more detailed analysis of the flavor constraints would be interesting.

\subsubsection{Trinification}
A final example consists of a similar generalization of trinification \cite{Glashow:1984gc}. Concretely, consider the breaking pattern
\be
SU(6)\times SU(6)\times SU(6)/\mathbb{Z}_2 \rightarrow [SU(3)\times SU(3)\times SU(3)]^2 \times U(1)^3
\ee
As in the Pati-Salam case, one copy of the trinification group will break to the Standard Model group, and the other to its twin. This again insures a private hypercharge for the Standard Model. In this a case full generation plus twin is contained in 
\be
( \six , \bar \six , \one)\oplus( \one , \six ,\bar \six) \oplus( \bar \six , \one, \six)
\ee
The trinification representations support a two Higgs doublet model in the bulk and both the Standard Model and the twin sector are again anomaly free by construction. A slightly unpleasant feature of this construction is that the lepton yukawa of the generation that resides in the bulk is forbidden by the $SU(6)\times SU(6)\times SU(6)$ gauge symmetry. This implies that this yukawa must arise from a an irrelevant operator involving some boundary-localized spurions that break the bulk gauge symmetry.
The argument for the absence of proton decay and neutron/anti-neutron oscillations is analogous to the argument given for the $U(4)\times U(6)$ model.

Note that both the trinification and Pati-Salam setup each contribute one dark neutrino to the hidden sector, as well as a dark sterile neutrino. To prevent the dark neutrino from contributing to $N_{eff}$ one would have to ensure that it is lifted by the sterile neutrino, sufficiently weakly coupled to fall out of thermal equilibrium at an early stage, or cooled relative to the Standard Model bath by preferential reheating in the Standard Model sector.

\section{Conclusions}
In this paper we have constructed a general class of models where the Higgs is a pseudo-goldstone boson in an orbifolded gauge theory. The key feature is the relation between a theory and its orbifold-daughters, which imbues the orbifold daughter with the precise couplings and discrete symmetries necessary to realize an accidental symmetry of the quadratic action and identify the Standard Model-like Higgs as a corresponding pseudo-goldstone. The components of an orbifold Higgs model that are relevant for naturalness are fully determined by the group theoretical properties the orbifold group. As such, we provide a complete classification of such `model fragments' consisting out of the Higgs and the top quarks: Every orbifold group corresponds to a unique quiver which is fully specified by the dimensions of the irreducible representations of the orbifold group. The twin Higgs model fits in this framework as the particular case of a $\mathbb{Z}_2$ orbifold. 
Besides providing a generalization of twin Higgs, a considerable advantage of the orbifold interpretation is a natural explanation for the accidental symmetries playing a crucial role; such symmetries (in particular $SU(4)$) would be an ad-hoc feature in a pure twin Higgs setup. While continuous symmetries of the quartic are also not guaranteed by the orbifold procedure, they provide a clear indication of the dynamics required for a viable UV completion (namely dominance of the double-trace quartic in the parent theory). 
Another bonus --related to the geometric interpretation-- is an elegant mechanism for reducing tension with cosmological observables, by localizing the light Standard Model degrees of freedom on the orbifold fixed points.

The reader may reasonably complain that generalizations beyond the abelian orbifolds are baroque and introduce numerous complications, including potential (modest) cutoff sensitivity at one loop and possible cosmological complications from the abundance of light non-abelian gauge bosons. While this may be so, the deeply valuable point is that they provide proof of principle that {\it partner states in natural theories of the weak scale may take an entirely unexplored form}. In the case of the $S_3$ orbifold this is particularly clear: the quadratic sensitivity of the top loop to higher scales is cancelled by a loop of Standard Model-neutral top partners with twice the multiplicity and half the coupling-squared.

There are a large number of possible future directions: 
\begin{itemize}
\item So far we only constructed a rough sketch of some UV completions in terms of higher dimensional geometric orbifolds and it would certainly be worthwhile to map out the possibilities in a more systematic fashion. In this work we for simplicity restricted ourselves to flat extra dimensions, however warped constructions should be possible as well. In fact, while this paper was in preparation, a first example of a holographic UV completion\footnote{Their setup is slightly different from ours, in the sense that they make use of a set of generic Dirichlet/Neumann boundary conditions to break the bulk symmetry, rather than of a particular orbifold projection.} of the twin Higgs already appeared \cite{Geller:2014kta}.

A particularly intriguing property of the examples we constructed so far is the requirement of low scale gauge coupling unification to successfully embed the hypercharge group, and it would be interesting to elucidate this apparent connection further. Of particular interest are also explicit geometric and deconstructed realizations of the non-abelian orbifolds. 

\item Our formalism is entirely transferable to supersymmetric theories, and a similar classification of orbifold supersymmetric models should be possible. The main difference in this case is the presence of an $R$-symmetry, which will participate in the orbifold projection. The simplest example of this mechanism is already known in the form of folded supersymmetry \cite{Burdman:2006tz}, but even more interesting possibilities include simultaneous orbifolds of the weak gauge group and an $R$-symmetry. In a different vein, it would be interesting to simply construct spontaneously broken supersymmetric UV completions of the orbifold Higgs, along the lines of \cite{Craig:2013fga}. 

\item Thus far we have restricted ourselves to the regular embeddings, however {\it a priori} there are many other representations to choose from. For arbitrary representations the necessary rescaling of the coupling constants is no longer guaranteed, however it is easy to verify that for certain sub-representations of the regular representation, the coupling constants do exhibit the correct scaling behavior. Moreover these alternative representations are expected to give rise to more exotic quivers than the ones discussed in this paper. Another direction would be to consider other geometric defects such as orientifolds that arise in string theory and possess field theory analogues.

\item Perhaps the most important opportunity lays in mapping out the detailed phenomenology of this class of models across the experimental frontiers, from colliders to dark matter to low-energy probes. The prime hallmark of orbifold Higgs models is a modification of the Higgs couplings through mixing with the Higgs(es) in the dark sectors, as is the case for the twin Higgs \cite{Craig:2013xia,Burdman:2014zta}. 

Further innovations include additional confining sectors connected through the Higgs portal \cite{Silveira:1985rk}, 
as well as additional pseudo-goldstone states near the weak scale that couple to the Standard Model primarily through irrelevant operators. A ramification of UV completions of the Orbifold Higgs (either in terms of extra dimensions or quiver gauge theories) is that the higher KK-modes may be accessible, either directly at a future high energy collider, or indirectly as portal between the Standard Model and the dark sectors.

 The LHC phenomenology would be similar to that of a hidden valley \cite{Strassler:2006im} through the Higgs portal \cite{Juknevich:2009gg,Raman}.
Note that explicit UV completions may also provide a handle via low-energy constraints not encountered at the level of the field theory orbifold. For example, in the models explored here the daughter states (corresponding to the untwisted sector of some geometric completion) introduce no flavor violation, while introducing heavier modes in the twisted sector may mediate flavor-violating processes at finite loop order. 

The final frontier awaiting exploration is that of dark matter. Since the dark sector and the Standard Model share baryon number, the Orbifold Higgs may naturally harbor an asymmetric dark matter candidate \cite{Kaplan:2009ag} in the form of the lightest dark baryon. Interestingly, the dark sector in an Orbifold Higgs model is generically a factor of a few heavier than the Standard Model sector, such that these models may automatically generate the right dark matter/baryon ratio in the universe.

\end{itemize}

\section*{Acknowledgements} We thank Aria Basirnia, Zackaria Chacko, Aleksey Cherman, Tony Gherghetta, Roni Harnik, Kiel Howe, Andrey Katz, Alberto Mariotti, Yasunori Nomura, Duccio Pappadopulo, Michele Papucci, Gilad Perez, Stuart Raby, Michael Ratz, Michele Redi, Martin Schmaltz, Raman Sundrum, and Jure Zupan for helpful discussions. We especially thank Matt Strassler both for particularly useful conversations and for asking the question, ``how do you generalize the Twin Higgs?'' This work was supported in part by DOE grants SC0010008, ARRA-SC0003883, and DE-SC0007897. N.C. acknowledges support from the Aspen Center for Physics and NSF grant 1066293 where this work was partially completed. This manuscript has been authored by an author (SK) at Lawrence Berkeley National Laboratory under Contract No. DE-AC02-05CH11231 with the U.S. Department of Energy. The U.S. Government retains, and the publisher, by accepting the article for publication, acknowledges, that the U.S. Government retains a non-exclusive, paid-up, irrevocable, world-wide license to publish or reproduce the published form of this manuscript, or allow others to do so, for U.S. Government purposes.

\appendix
\section{Another nonabelian orbifold: $A_4$}\label{app:A4-toy}
For completeness, in this Appendix we work out in detail the $A_{4}$ orbifold of a Yang-Mills-Higgs model. The story runs parallel to the $S_{3}$ case of section \ref{sec:S3-toy}.
$A_{4}$ is the subgroup of even permutations of $S_{4}$; it is of order 12 and is generated by two elements
\be
\begin{split}
	& a=(12)(34) \qquad \text{and}  %
	\qquad b=(123)\,.%
\end{split}
\ee
There are four conjugacy classes
\be
	\CC_{0}=\{e\},\qquad \CC_{1}=\{b,ab,aba,ba \},\qquad \CC_{2} = \{ b^{2},ab^{2},b^{2}a,a b^{2}a \} ,\qquad \CC_{3} = \{ a,b^{2}ab,bab^{2}\}
\ee
we label group elements by an integer $s$, starting with $s=0$ for $e$, then $s=1,\dots,4$ for elements of $\CC_{1}$ in the order displayed above, and so on.

The regular representation has dimension 12. The decomposition works out easily: the trivial irrep has dimension $d_{0}=1$, so $12=1+d_{1}^{2}+d_{2}^{2}+d_{3}^{2}$ which entails $d_{1}=d_{2}=1$, $d_{3}=3$ (up to a permutation). Correspondingly there are four irreps
\be
\begin{array}{ccl}
	r_{0} \,:& \quad d_{0} =1 \quad &\quad  \text{trivial irrep} \\
	r_{1} \,:& \quad d_{1} =1 \quad &\quad  \text{1-dim'l irrep} \\
	r_{2} \,:& \quad d_{2} =1 \quad &\quad  \text{1-dim'l irrep} \\
	r_{3} \,:& \quad d_{3} =3 \quad &\quad  3\times 3 \\
\end{array}
\ee
whose matrix elements can be explicitly taken to be
\be
\begin{split}
	& r_{0}\,:\quad r_{0}^{s}=1\quad\forall s \\
	& r_{1}\,: \quad r_{1}^{s}=\left\{ \begin{array}{cl} %
	1 & g_{s}\in \CC_{0} \\ %
	\zeta & g_{s}\in \CC_{1}  \\%
	\zeta^{2} & g_{s}\in \CC_{2} \\ %
	1 & g_{s}\in\CC_{3} 
	\end{array}\right.
	\qquad \qquad 
	r_{2}\,: \quad r_{2}^{s}=\left\{ \begin{array}{cl} %
	1 & g_{s}\in \CC_{0} \\ %
	\zeta^{2} & g_{s}\in \CC_{1}  \\%
	\zeta^{} & g_{s}\in \CC_{2} \\ %
	1 & g_{s}\in\CC_{3} 
	\end{array}\right.\\
	& r_{3}\,: \quad   %
	r_{3}^{9}=r_{3}(a)=\left(\begin{array}{ccc} -1 & 0  &0  \\ 0 & 1 &0 \\0 & 0& -1 \end{array}\right) \qquad%
	r_{3}^{1}=r_{3}(b)=\left(\begin{array}{ccc} 0& 1 &0 \\ 0 &0 &1 \\ 1& 0& 0 \end{array}\right) 
\end{split}
\ee
where $\zeta=e^{2\pi i/3}$.

The \emph{regular} representation thus decomposes as
\be
	\gamma^{s}=r^{s}_{0}\oplus r^{s}_{1}\oplus r^{s}_{2}\oplus 3\cdot r^{s}_{3}
\ee

The mother theory is once again summarized by table (\ref{tab:mothertheory}), where now 
$\Gamma=12$ is fixed.
From the general discussion of section \ref{sec:general-features}, the pattern of symmetry breaking is
\be
\begin{split}
	SU(24) &\quad\longrightarrow \quad SU(2)_{0}\,\times \, SU(2)_{1}\,\times \, SU(2)_{2}\,\times \, SU(6)_{3} \,,\\
	SU(36) &\quad\longrightarrow \quad SU(3)_{0}\,\times \, SU(3)_{1}\,\times \, SU(3)_{2}\,\times \, SU(9)_{3} \,,\\
	SU(12) &\quad\longrightarrow \quad SU(3)_{3}\,.
\end{split}
\ee

Multi-indices of $SU(24)$ have therefore the following span:
\be
	A=(\alpha,a,\a):\quad \alpha=0,1,2,3\quad %
	a=\left\{\begin{array}{l}  %
	1,\dots,3\ \ \text{for} \ \alpha=0,1,2 \\%
	1,\dots,9\ \ \text{for} \ \alpha=3 %
	\end{array}\right.\quad %
	\a=\left\{\begin{array}{l}  %
	0\ \ \text{for} \  \alpha=0,1,2 \\%
	0,1,2\ \ \text{for} \ \alpha=3 %
	\end{array}\right.%
\ee
and so on for other symmetry groups.

\noindent The gluons $({A^{(\alpha),i}})_{A}^{\phantom{A}B}$ surviving the orbifold are the block-diagonal ones
\be
\begin{split}
	SU(N)_{0} &:\quad ({A^{(0),i_{0}}})_{(0,a,0)}^{\phantom{(0,a,0)}{(0,b,0)}}\qquad a,b=1,\dots,N \quad i_{0}=1,\dots,N^{2}-1\\
	SU(N)_{1} &:\quad ({A^{(1),i_{1}}})_{(1,a,0)}^{\phantom{(0,a,0)}{(1,b,0)}}\qquad a,b=1,\dots,N \quad i_{1}=1,\dots,N^{2}-1\\
	SU(N)_{2} &:\quad ({A^{(2),i_{2}}})_{(2,a,0)}^{\phantom{(0,a,0)}{(2,b,0)}}\qquad a,b=1,\dots,N \quad i_{2}=1,\dots,N^{2}-1\\
	SU(3N)_{3} &:\quad ({A^{(3),i_{3}}})_{(3,a,\a)}^{\phantom{(0,a,0)}{(3,b,\b)}}\qquad a,b=1,\dots,3N\quad \a,\b=0,1,2 \quad i_{3}=1,\dots,9N^{2}-1
\end{split}
\ee
where Schur's lemma requires that the gluons of $SU(3N)_{3}$ be of the form
\be
	({A^{(3),i}})_{(3,a,\a)}^{\phantom{(0,a,0)}{(3,b,\b)}} = ({A^{(3),i}})_{(3,a)}^{\phantom{(0,a}{(3,b)}}\,\delta_{\a}^{\b}\,.
\ee

On the other hand, the $\G$-action on the $SU(12)$-flavor fundamentals reads simply
\be
\begin{split}
	& (\gamma_{F}^{s})_{M}^{\phantom{M}N} = (\one_{F}\otimes \gamma_{}^{s})_{M}^{\phantom{M}N} = \delta_{\mu}^{\phantom{\mu}\nu}\delta_{m}^{\phantom{m}n}\, (r^{s}_{\mu})_{\m}^{\phantom{\m}\n} 
	\qquad \text{with }\ 	M = (\mu,m,\m)\\
	& \mu=0,1,2,3 \quad %
	m=\left\{\begin{array}{l}  %
	1\ \ \text{for} \ \mu=0,1,2 \\%
	1,2,3\ \ \text{for} \ \mu=3 %
	\end{array}\right.\quad %
	\m=\left\{\begin{array}{l}  %
	0\ \ \text{for} \  \mu=0,1,2 \\%
	0,1,2\ \ \text{for} \ \mu=3 %
	\end{array}\right.%
\end{split}
\ee

Turning to the matter fields, we employ once again formula (\ref{eq:bifund-invt}). For example, the transformation of $H$ under $g_{s}\in\G$ then reads $H_{A}^{\phantom{A}M}\,\mapsto\, (\gamma^{s}_{N})_{A}^{\phantom{A} B}\ (\gamma^{s\,\,*}_{F})_{\phantom{A} N}^{M}\ H_{B}^{\phantom{A}N}$, and all we need to do is examine the invariant fields case by case, i.e. for $\alpha=0,1,2,3$.

\noindent{\underline{$\alpha=\mu=0$}}
\be
	d_{\alpha}=d_{\mu}=1,\qquad \a,\m=0, \quad a=1,2,\quad m=1
\ee
so we are left with
\be
	(h^{(0)})_{a}\,=\,H_{(0,a,0)}^{\phantom{(0,m,0)}(0,m,0)}
\ee
in the $\Box\ \text{of}\ SU(2)_{0}$.

\noindent{\underline{$\alpha=1,\,\mu=2$}}
\be
\begin{split}
	& d_{\alpha}=d_{\mu}=1,\qquad \a,\m=0, \quad a=1,2,\quad m=1 
\end{split}
\ee
so we are left with
\be
	(h^{(1)})_{a}\,=\,H_{(1,a,0)}^{\phantom{(0,m,0)}(2,m,0)}
\ee
in the $\Box\ \text{of}\ SU(2)_{1}$.

\noindent{\underline{$\alpha=2,\,\mu=1$}}\\
This is very similar to the previous case, we find the invariant combination
\be
	(h^{(2)})_{a}\,=\,H_{(2,a,0)}^{\phantom{(0,m,0)}(1,m,0)}
\ee
in the $\Box\ \text{of}\ SU(2)_{2}$.

\noindent{\underline{$\alpha=\mu=3$}}
\be
\begin{split}
	d_{\alpha}=d_{\mu}=3,\quad &\a,\m=0,1,2, \quad a=1,\dots,6,\quad m=1,2,3 
\end{split}
\ee
the invariant combination is then
\be
\begin{split}		
	& (h^{(3)})_{a}^{\phantom{a}m}\,=\,{1\over \sqrt{3}}\left( H_{(3,a,0)}^{\phantom{(0,m,0)}(3,m,0)}+ H_{(3,a,1)}^{\phantom{(0,m,0)}(3,m,1)}+ H_{(3,a,2)}^{\phantom{(0,m,0)}(3,m,2)}\right) 
\end{split}
\ee
in the $\Box\times\overline\Box\ \text{of}\ SU(6)_{3}\times SU(3)_{3}$.

Overall we found the following Higgs content:
\be
\resizebox{0.85\hsize}{!}{$\displaystyle 
\begin{array}{c|c|cccc|c}
	& \text{mother theory d.o.f.} & SU(2)_{0} & SU(2)_{1} & SU(2)_{2} & SU(6)_{3} & SU(3)_{3} \\
	\hline%
	h^{(0)} & H_{(0,a,0)}^{\phantom{(0,m,0)}(0,m,0)} & \Box & 1 & 1&1 & 1 \\
	h^{(1)} & H_{(1,a,0)}^{\phantom{(0,m,0)}(2,m,0)}  & 1 & \Box & 1 & 1 &1 \\
	h^{(2)} & H_{(2,a,0)}^{\phantom{(0,m,0)}(1,m,0)}  & 1 & 1 & \Box & 1 & 1\\
	h^{(3)} & {1\over \sqrt{3}}\left( H_{(3,a,0)}^{\phantom{(0,m,0)}(3,m,0)}+ H_{(3,a,1)}^{\phantom{(0,m,0)}(3,m,1)}+ H_{(3,a,2)}^{\phantom{(0,m,0)}(3,m,2)} \right)  & 1 & 1 & 1 & \Box &  \overline\Box 
\end{array}
$}
\ee

As a partial check, 
we may employ the Clebsch-Gordan decomposition of $A_{4}$ irreps (which can easily be worked out by standard representation theory, or found in the math literature):
\be\label{eq:CG-table-A4}
\begin{split}
 r_0\otimes r_\alpha & = r_\alpha \qquad \alpha=0,1,2,3\\
 r_1\otimes r_1 &  =r_2 \qquad %
 r_1\otimes r_2  =r_0 \qquad  %
 r_1\otimes r_3  = r_3 \\
 r_2\otimes r_2 & = r_1 \qquad %
 r_2\otimes r_3  =r_3 \\
 r_3\otimes r_3 & = r_0\oplus r_1\oplus r_2\oplus  \mathbb{1}_{2}\otimes r_3 \,,\\
\end{split}
\ee
to study the tensor product of gauge and flavor $\G$-representations. This is easily seen to decompose into
\be
\begin{split}
	\gamma_{N}\otimes \gamma_{F} & =\mathbb{1}_{N\cdot F}\otimes \gamma\otimes\gamma \\
	& = \mathbb{1}_{N\cdot F}\otimes \big( \mathbb{1}_{12}\otimes r_{0}  \oplus \mathbb{1}_{12} \otimes r_{1} \oplus \mathbb{1}_{12} \otimes r_{2}\oplus \mathbb{1}_{36} \otimes r_{3} \big)
\end{split}
\ee
meaning that the orbifold preserves $12\, N\cdot F = 24$ Higgs components, in agreement with the above result.
The same analysis carries over to $Q$ and $U$, we omit the explicit tables since the generalization is fairly obvious.

\begin{figure}[h!]
\begin{center}
\includegraphics[width=0.8\textwidth]{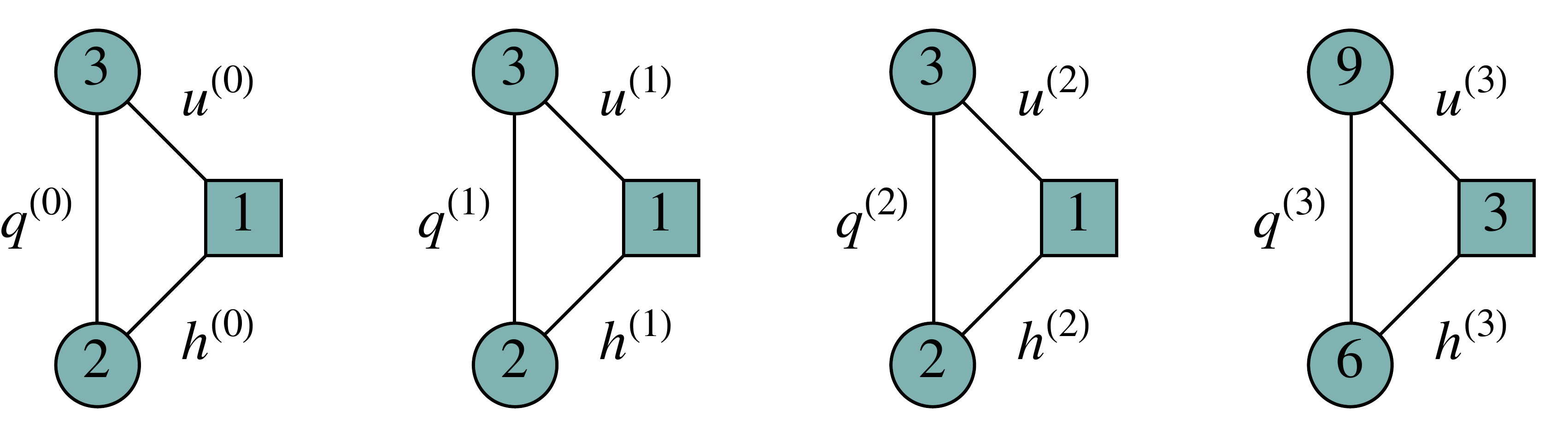} 
\caption{The quiver of the $A_{4}$ orbifold.}
\label{fig:A4quiver}
\end{center}
\end{figure}

In this case we find a daughter theory with $4$ sectors: as for $\G=S_{3}$, each sector has gauge symetry  $SU(2 \, d_{\alpha})^{(\alpha)}\times SU(3\, d_{\alpha})^{(\alpha)}$, together with matter consisting of $h^{(\alpha)},\, q^{(\alpha)},\, u^{(\alpha)}$ in the corresponding (bi)-fundamental representations. The stucture of this daughter theory is conveniently summarized by the quiver diagram of figure \ref{fig:A4quiver}.

Once again, the general discussion of section \ref{sec:scaling} implies that Yukawa, gauge, and quartic couplings in the mother theory generate corresponding interactions in the daughter theory, privately within each sector. This example features a nontrivial rescaling of the couplings, for the sector corresponding to the $3$-dimensional representation of $A_{4}$.

The quiver structure exhibits manifestly the $S_{3}$ symmetry of the daughter theory, which acts by permuting the $\alpha=0,1,2$ sectors. Note however that the {approximate continuous symmetry} protecting the Higgs mass is now an SU(24) rotating the $24$ surviving Higgs components.

\providecommand{\href}[2]{#2}\begingroup\raggedright\endgroup



\end{document}